\documentclass[conference]{IEEEtran}
\IEEEoverridecommandlockouts
\usepackage{cite}
\usepackage{amsmath,amssymb,amsfonts}
\usepackage[noend]{algorithmic}
\usepackage{graphicx}
\usepackage{textcomp}
\usepackage{xcolor}
\def\BibTeX{{\rm B\kern-.05em{\sc i\kern-.025em b}\kern-.08em
    T\kern-.1667em\lower.7ex\hbox{E}\kern-.125emX}}
\usepackage{amsmath}
\usepackage{amssymb}
\usepackage{amsthm}
\usepackage{mathabx}
\usepackage{esint}
\usepackage{xcolor}
\usepackage{caption}
\usepackage{subcaption} 
\usepackage{hyperref}
\usepackage{longtable,tabularx}
\usepackage{todonotes}
\usepackage{tabularx}
\usepackage{booktabs}
\usepackage{enumitem}

\newtheorem{probl}{Problem}

\newcommand{\vMinj}{{v}_\mathit{\min, j}} 
\newcommand{\vMaxj}{{v}_\mathit{\max, j}} 
 
\newcommand{\aEmergency}{{a}_\mathit{\max}} 
 
\newcommand{\bEmergency}{{b}_\mathit{\min}}

\newcommand{\CWP}[1]{\mathit{CWP}_{#1}}

\usepackage{multirow}
\usepackage{graphicx}
\usepackage{epstopdf}
\usepackage{algorithm}
\usepackage{algorithmic}

\begin{document}

\title{ETA Coordination at UAM Corridor Merging Points Using Worst-Case and Stochastic Trajectory Bounds
% \\
% {\footnotesize \textsuperscript{*}Note: Sub-titles are not captured for https://ieeexplore.ieee.org  and
}

% \and
% \and
% \and
% \and
% \and
\author{
\IEEEauthorblockN{Sasinee Pruekprasert}
\IEEEauthorblockA{
\textit{The University of Tokyo},\\Tokyo, Japan\\
spruekprasert@g.ecc.u-tokyo.ac.jp
} 
\and
\IEEEauthorblockN{Shinji Nakadai}
\IEEEauthorblockA{
\textit{Intent Exchange, Inc.}\\
Tokyo, Japan\\
nakadai@intent-exchange.com
}\and
\IEEEauthorblockN{Katsuhiro Nishinari}
\IEEEauthorblockA{
\textit{The University of Tokyo}\\
Tokyo, Japan\\
tknishi@mail.ecc.u-tokyo.ac.jp
}
}

\author{
\IEEEauthorblockN{Sasinee Pruekprasert}
\IEEEauthorblockA{
\textit{The University of Tokyo},\\Tokyo, Japan\\
spruekprasert@g.ecc.u-tokyo.ac.jp
} 
\and
\IEEEauthorblockN{Shinji Nakadai}
\IEEEauthorblockA{
\textit{Intent Exchange, Inc.}\\
Tokyo, Japan\\
nakadai@intent-exchange.com
}\and
\IEEEauthorblockN{Katsuhiro Nishinari}
\IEEEauthorblockA{
\textit{The University of Tokyo}\\
Tokyo, Japan\\
tknishi@mail.ecc.u-tokyo.ac.jp
}
}

\thispagestyle{empty}
\onecolumn
{\large
\noindent 
This is a preprint of a paper accepted for publication in the proceedings of the 45th Digital Avionics Systems Conference (DASC 2026).

\vspace{1em}

\noindent \copyright\ 2026 IEEE. Personal use of this material is permitted. Permission from IEEE must be obtained for all other uses, in any current or future media, including reprinting/republishing this material for advertising or promotional purposes, creating new collective works, for resale or redistribution to servers or lists, or reuse of any copyrighted component of this work in other works. 
}
\newpage
\twocolumn
\setcounter{page}{1}

\maketitle

\begin{abstract}
We study an Estimated Time of Arrival (ETA)-based traffic-coordination framework for Urban Air Mobility corridors with merging at constrained waypoints (CWPs), where approved ETAs at CWPs serve as Required Times of Arrival (RTAs). Vehicle operators submit ETA plans at the merging point for approval by corridor-management authorities before corridor entry. Corridor entry is then scheduled by enforcing pairwise ETA gaps that maintain inter-vehicle separation on shared corridor sections. We develop two trajectory bounds to compute sufficient ETA gaps: a worst-case bound based on prescribed speed limits, and a stochastic bound based on probabilistic position envelopes under acceleration uncertainty. Using these bounds, we formulate sufficient ETA-gap computation and first-come, first-served corridor entrance scheduling. Simulations show that ETA coordination improves safety over an unscheduled baseline. The worst-case bound provides stronger robustness under higher disturbance levels, whereas the stochastic bound allows higher throughput under mild disturbances while relying on probabilistic modeling assumptions.
\end{abstract}

\begin{IEEEkeywords}
UAM corridors, Arrival scheduling, Urban Air Mobility, Air traffic management
\end{IEEEkeywords}

\section{Introduction}
\label{sec:introduction}

Urban Air Mobility (UAM) is being considered as a low-altitude mode of transportation for passengers and cargo in metropolitan areas, supported by advances in electric vertical take-off and landing (eVTOL) aircraft, automation, and on-demand mobility concepts~\cite{mueller2017enabling,keeler2019investigation,verma2022design}.
While such operations may help reduce ground congestion and expand transportation options, they also introduce airspace-management challenges, including traffic complexity, infrastructure needs, controller workload, interaction with conventional aerial vehicle operations, and community concerns such as noise~\cite{keeler2019investigation,gao2024noise}.
As UAM traffic increases, scalable coordination methods will be needed to maintain safety and efficiency without adding excessive burden to the current Air Traffic Control (ATC) system.

One approach for organizing dense UAM traffic is to use structured low-altitude routes or corridors~\cite{fontaine2023urban,verma2022design}.
These corridors define reserved airspace volumes connecting important urban locations and can support traffic management, separation assurance, and compatibility with surrounding airspace.
In corridor-based UAM concepts of operation, traffic may be coordinated through service entities such as Providers of Services for UAM (PSUs), under regulatory oversight and alongside the broader ATC system~\cite{fontaine2023urban, keeler2019investigation,verma2022design}.
Compared with less-structured airspace, corridor-based operations reduce the freedom of individual vehicles, but can simplify coordination and make high-density traffic more manageable~\cite{bauranov2021designing}.

Several studies have examined the design and evaluation of UAM and Unmanned Aerial Vehicle (UAV) corridor systems.
Muna et al. proposed a corridor framework based on air cubes, skylanes, intersections, vertiports, and gates, and analyzed capacity and congestion-related properties of such structures~\cite{muna2021air}.
Jiang et al. introduced evaluation metrics for corridor designs using traffic data together with spatial and temporal indicators related to safety and environmental impact~\cite{jiang2022metrics}.
Other studies have considered corridor-network design, congestion mitigation, and architecture-aware planning for UAM systems~\cite{wang2021air,weigang2025eixao}.
These studies suggest that corridor-based operations can play an important role in improving the safety, efficiency, and capacity of UAM operations.
 
In addition to UAM corridor design, practical operations also require scheduling methods that regulate when vehicles enter and traverse shared airspace.
Previous research has considered scheduling and resource-allocation problems in related Advanced Air Mobility (AAM) and UAV settings, including dynamic task scheduling, demand--capacity balancing, vertiport allocation, and simulation-based scheduling~\cite{halder2022dynamic,yokoyama2025performance,weigang2025eixao}.
More directly related to merging and intersection operations, Balachandran et al. proposed a distributed consensus framework in which UAS approaching a common intersection exchange arrival-time intervals and independently compute safe crossing times using synchronized information~\cite{balachandran2018distributed}.
Yahi et al. addressed conflict detection and resolution for UAM aircraft at intersections, using a decentralized communication architecture and a receding-horizon trajectory planner based on nonlinear model predictive control to generate conflict-resolution maneuvers~\cite{yahi2024receding}.
Liu et al. studied take-off and merging control in UAM corridors by integrating tactical take-off-time coordination and dynamic merging-point selection with operational trajectory optimization~\cite{liu2026integrated}.
These studies demonstrate the importance of coordination at UAM intersections and merging points.
However, they primarily coordinate merging or intersection traffic through consensus, conflict-resolution maneuvers, or trajectory optimization, rather than deriving analytical ETA-based separation requirements for corridor-access scheduling.

Estimated Time of Arrival (ETA)-based coordination provides a lightweight way to regulate corridor access using timing information at constrained waypoints (CWPs).
Previous ETA-based studies have considered safety-aware scheduling for a single corridor section~\cite{pruekprasert2025safe,pruekprasert2024enhancing} and for sequential corridor sections~\cite{fujita2026visual,pruekprasert2026ifac}.
These studies show that ETA gaps can be used as scheduling variables for maintaining separation under corridor-level constraints.
In merging corridors, however, vehicles may enter from different upstream branches and share only part of their routes.
Therefore, the required ETA gap must depend on the corridor sections actually shared by each consecutive vehicle pair.

% This paper studies an ETA-based coordination framework for merging operations in UAM corridors.
% Under this framework, vehicle operators submit ETA requests at CWPs to corridor-management authorities, such as PSUs, which review the requests and, when necessary, adjust them before authorizing corridor entry.
% Using CWP-level ETA information rather than full within-corridor trajectory plans, the framework imposes high-level restrictions on corridor access and progression, while real-time trajectory tracking and collision avoidance remain the responsibility of individual vehicles.

This paper studies an ETA-based coordination framework for merging operations in UAM corridors.
Under this framework, vehicle operators submit ETA requests at CWPs to corridor-management authorities, such as PSUs, which review the requests and, when necessary, adjust them before authorizing corridor entry.
Once approved, these ETAs play a role similar to Required Times of Arrival (RTAs), since they specify CWP arrival times that vehicles are expected to satisfy.
Using CWP-level ETA information rather than full within-corridor trajectory plans, the framework imposes high-level restrictions on corridor access and progression, while real-time trajectory tracking and collision avoidance remain the responsibility of individual vehicles.

A key contribution of the proposed approach is the derivation of analytical ETA-based separation requirements for merging corridors from the corridor sections shared by consecutive vehicle pairs.
This provides a lightweight corridor-access scheduling criterion that complements existing consensus-based, conflict-resolution-based, and trajectory-optimization-based approaches.
We develop two types of trajectory bounds: worst-case bounds, adapted from previous ETA-gap constructions for corridor scheduling~\cite{pruekprasert2025safe,pruekprasert2026ifac}, and stochastic bounds derived from linear-Gaussian uncertainty propagation, terminal conditioning, and smoothing.
Based on these bounds, we formulate sufficient ETA-gap computations for consecutive vehicles sharing one or more corridor sections, and develop a first-come, first-served corridor entrance-scheduling algorithm to coordinate corridor access.

% The rest of the paper is organized as follows. 
% Section~\ref{sec:eta_coordination} introduces the ETA coordination framework, the merging corridor model, and the safe ETA-gap problem. 
% Sections~\ref{sec:wc_eta_gap} and~\ref{sec:stoc_eta_gap} present the worst-case and stochastic trajectory bounds for ETA-gap computation. 
% Section~\ref{sec:entrance_scheduling} describes first-come, first-served corridor entrance scheduling based on the computed gaps. 
% Section~\ref{sec:simulation} reports the numerical simulation results. Section~\ref{sec:conclusion} concludes the paper.
The rest of the paper is organized as follows. 
Section~\ref{sec:eta_coordination} presents the ETA coordination framework, merging corridor model, and safe ETA-gap problem. 
Sections~\ref{sec:wc_eta_gap}--\ref{sec:entrance_scheduling} develop the trajectory bounds and scheduling algorithm. 
Section~\ref{sec:simulation} gives simulation results, and Section~\ref{sec:conclusion} concludes the paper.

\section{ETA Coordination in Merging UAM Corridors}
\label{sec:eta_coordination}
This section introduces the Estimated Time of Arrival (ETA)-based coordination framework and the merging corridor model studied in this work. We then formulate the safe ETA-gap problem at the merging point, which serves as the basis for the safety-bound computations in the following sections.

\begin{figure}[tbp]
\centerline{\includegraphics[width=0.9\columnwidth]{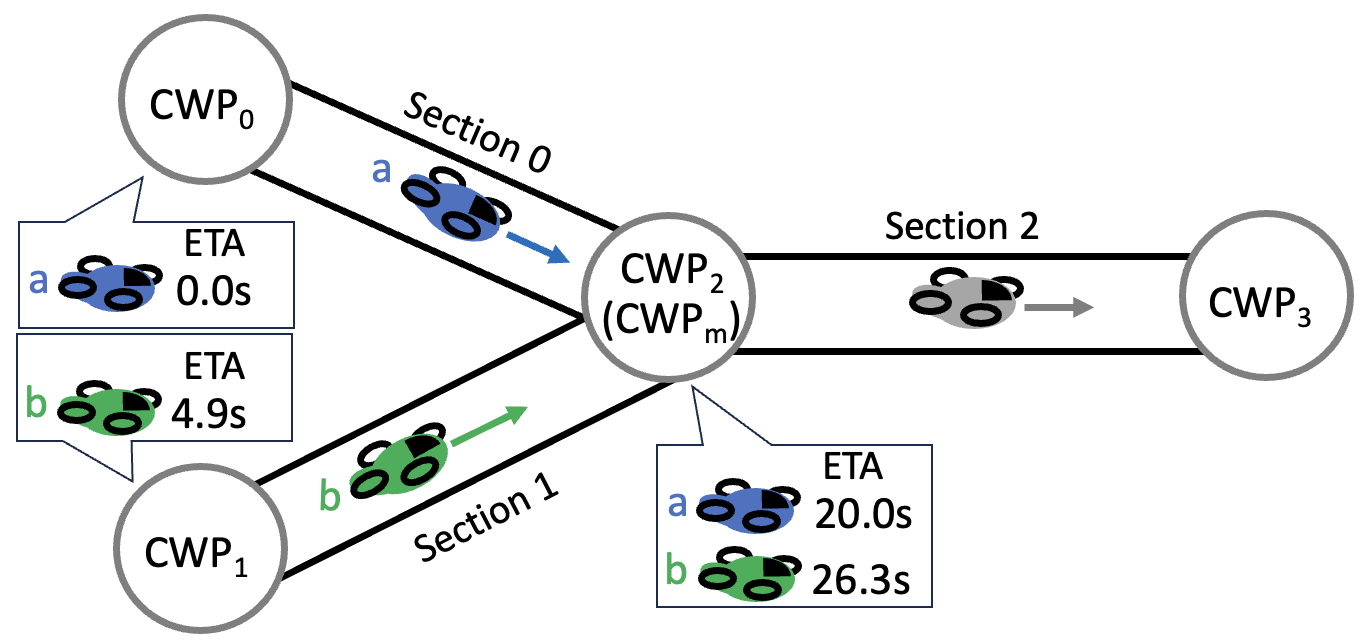}}
\caption{Merging UAM corridor network with CWP-based ETA coordination. Vehicles enter at either $\CWP{0}$ or $\CWP{1}$, merge at $\CWP{2}$, and proceed to $\CWP{3}$. 
%In this example, the ETA gap at $\CWP{2}$ is $\Delta t^{2}_{(a,b)} = 6.3$ s.
}
\label{fig:UAM_merge}
\end{figure}

\subsection{Coordination Framework}

We consider ETA-based air traffic coordination for maintaining safety-assured inter-vehicle separation in UAM corridors, motivated by emerging AAM operations in dense urban environments.
UAM corridors are reserved airspace volumes whose access and operational constraints are managed by designated authorities, such as Providers of Services for UAM (PSUs).
The proposed approach uses Estimated Times of Arrival (ETAs) at Constrained Waypoints (CWPs) as the coordination variables.
As illustrated in Fig.~\ref{fig:UAM_merge}, operators of the vehicles submit proposed ETAs at CWPs to a PSU, which validates and, when necessary, adjusts them before authorizing vehicles to enter constrained corridor sections.
Rather than requiring complete within-corridor flight plans, our framework relies on CWP-level ETA information, reducing the amount of coordination data.

We focus on ETA scheduling before vehicles enter constrained corridor sections, where the order at the merging point is determined on a first-come, first-served basis. The ETA schedule approved by the PSU provides a network-level safety restriction based on the ETA information exchanged with vehicle operators. Real-time trajectory tracking and collision-avoidance actions remain the responsibility of the individual vehicles, which execute these functions locally through autonomous operation or human piloting. Thus, the proposed ETA coordination framework is intended to constrain corridor entry and progression at the scheduling level, while leaving low-level trajectory execution to the vehicles.

\subsection{Merging Corridor Model} \label{subsec:merging}

We consider a UAM corridor network consisting of a finite number of corridor sections connected through CWPs. In the merging scenario illustrated in Fig.~\ref{fig:UAM_merge}, vehicles may enter the network from either $\CWP{0}$ or $\CWP{1}$ and then merge at $\CWP{2}$. The vehicles are indexed by $i \in \{0,\dots,n-1\}$, where the index denotes the order in which they are scheduled to enter the shared downstream corridor after the merge.

Within the corridor network, vehicle motion is represented along the centerline of each corridor section by a one-dimensional kinematic model following~\cite{pruekprasert2025safe,fujita2026visual},
\begin{equation}
\label{eq:kinematic}
    \dot{x}_i(t) = v_i(t), \qquad 
    \dot{v}_i(t) = a_i(t),
\end{equation}
where $x_i(t)$ is the along-section position of vehicle~$i$, $v_i(t)$ is its speed, and $a_i(t)$ is its acceleration input at time $t$.

Each corridor section~$j$ has admissible speed limits $\vMinj$ and $\vMaxj$, where $\vMinj<\vMaxj$, determined by corridor design, operational requirements, or airspace regulations. When vehicle~$i$ is traveling in section~$j$, its speed is required to satisfy
\begin{equation}
\label{eq:vlim}
    0 < \vMinj \le v_i(t) \le \vMaxj .
\end{equation}

% For scheduling purposes, we assign a nominal traversal time to each section~$j$ as
% $ 
%     \tau_j = \frac{l_j}{v_{\mathit{avg},j}}$, where $ 
%     v_{\mathit{avg},j} = \frac{\vMinj + \vMaxj}{2}$,
% and $l_j$ is the length of section~$j$. This common section-wise travel time is used to simplify the ETA scheduling problem, although the formulation can be extended to vehicle-dependent or route-dependent traversal times.

We assign a nominal traversal time to each section~$j$ as
\begin{equation}
\label{eq:tau}
    \tau_j = \frac{l_j}{v_{\mathit{avg},j}},
    \qquad
    v_{\mathit{avg},j} = \frac{\vMinj+\vMaxj}{2},
\end{equation}
where $l_j>0$ is the length of corridor section~$j$.
This common section-wise travel time is used to simplify the ETA scheduling problem, although the formulation can be extended to vehicle-dependent or route-dependent traversal times.

Moreover, each vehicle is assumed to operate under bounded acceleration and braking:
\begin{equation}
\label{eq:acc_bounds}
    \bEmergency \le a_i(t) \le \aEmergency,
\end{equation}
where $\aEmergency \ge 0$ denotes the upper acceleration limit, and
$\bEmergency \le 0$ denotes the lower acceleration limit corresponding to maximum braking.

\subsection{Safe ETA-Gap Problem at the Merging Point}
\label{sec:problem_formulation}

We adopt a fixed minimum separation distance based on Near Mid-Air Collision (NMAC) avoidance rules as the safety requirement for ETA-based scheduling. NMAC rules specify a minimum distance that should be preserved between aerial vehicles to avoid unsafe proximity~\cite{johnson2017exploration, muna2021air}. 
% This assumption follows distance-based safety practices in conventional aviation, where maintaining a prescribed separation threshold during nominal operation provides a safety margin for vehicle-level avoidance actions under disturbances, tracking errors, or emergency maneuvers.
This assumption is consistent with distance-based safety practice in aviation, where maintaining a separation threshold provides margin for local avoidance actions under off-nominal conditions.

Let $d_{\mathrm{safe}}$ denote the required inter-vehicle separation. The ETAs at CWPs must be approved by the PSU in such a way that any two vehicles sharing a corridor section remain separated by at least $d_{\mathrm{safe}}$ throughout their shared traversal.

We consider a UAM corridor network consisting of a finite number of upstream corridor sections that merge into a shared downstream section. Let $\CWP{m}$ denote the merging CWP. The example in Fig.~\ref{fig:UAM_merge} shows the case of two incoming sections merging at $\CWP{2}$, i.e., $m=2$, but the formulation below applies to a general merging point. Vehicles may enter the network through different upstream branches and are scheduled at $\CWP{m}$ in a first-come, first-served manner.

% Let $r_i$ denote the route of vehicle~$i$, which can be represented by its entry CWP and 
% Let $t_{i,j}$ be the ETA of vehicle~$i$ at $\CWP{j}$ whenever $\CWP{j}$ belongs to its route. For two vehicles $i$ and $i+1$ that both pass through $\CWP{j}$, we denote their ETA gap at $\CWP{j}$ by
% Let $c_i$ denote the entry CWP of vehicle~$i$. 
% For example, in Fig.~\ref{fig:UAM_merge}, $c_a=\CWP{0}$ and $c_b=\CWP{1}$. 
% By a slight abuse of notation, we also use $c_i$ to denote the index of the entry CWP when no confusion arises. 
% Thus, in Fig.~\ref{fig:UAM_merge}, we may write $c_a=0$ and $c_b=1$.
Let $c_i$ denote the entry CWP of vehicle~$i$. 
For example, in Fig.~\ref{fig:UAM_merge}, $c_a=\CWP{0}$ and $c_b=\CWP{1}$. 
When convenient, we also use $c_i$ to denote the index of the entry CWP. 
Thus, in Fig.~\ref{fig:UAM_merge}, we may use $c_a$ for 0 and $c_b$ for 1.
Let $t_{i,j}$ denote the ETA of vehicle~$i$ at $\CWP{j}$ whenever its route passes through $\CWP{j}$. 
For two vehicles $i$ and $i+1$ that both pass through $\CWP{j}$, we denote their ETA gap at $\CWP{j}$ by
\[
    \Delta t^{j}_{(i,i+1)}
    :=
    t_{i+1,j} - t_{i,j}.
\]
In particular, $t_{i,m}$ denotes the ETA of vehicle~$i$ at the merging point $\CWP{m}$, and $\Delta t^{m}_{(i,i+1)}$ denotes the ETA gap between two consecutive vehicles at $\CWP{m}$.

For two vehicles $i$ and $i+1$ that are consecutive in the scheduled order at $\CWP{m}$, the merging ETA gap is
\[
    \Delta t^{m}_{(i,i+1)}
    =
    t_{i+1,m} - t_{i,m} > 0 .
\]
The problem is to determine a sufficiently small value of $\Delta t^{m}_{(i,i+1)}$ that guarantees safe separation over all corridor sections shared by the two vehicles.

\begin{probl}
\label{probl:etagap_merge}
For two vehicles $i$ and $i+1$ scheduled consecutively at the merging point $\CWP{m}$, find
\[
    \Delta t^{m}_{(i,i+1)}
    :=
    t_{i+1,m} - t_{i,m} > 0
\]
such that
\begin{align}\label{eq distance}
    x_i(t) - x_{i+1}(t)
    \ge d_{\mathrm{safe}}
\end{align}
for all times $t$ during which both vehicles occupy a shared corridor section, under the assumed vehicle dynamics and uncertainty model. Here, $x_i(t)$ and $x_{i+1}(t)$ denote the along-corridor positions of vehicles $i$ and $i+1$ measured on the shared section under consideration.
\end{probl}

In this formulation, the primary scheduling variable is the ETA difference at the merging CWP. %rather than an entry-time difference at a common upstream CWP. 
For vehicles entering from different branches, their upstream ETAs depend on their respective routes and section travel times. Once $\Delta t^{m}_{(i,i+1)}$ is determined at the merging point, the corresponding ETA gaps at upstream CWPs can be obtained by shifting this value according to the travel times along each incoming branch. 
% For example, in Fig.~\ref{fig:UAM_merge}, where $m=2$, the entry-time difference between vehicles entering from $\CWP{0}$ and $\CWP{1}$ can be derived from their desired arrival-time difference at $\CWP{2}$ and the traversal times from the respective entry CWPs to the merging point at $\CWP{2}$.

% The set of corridor sections shared by two consecutive vehicles depends on their routes. In Fig.~\ref{fig:UAM_merge}, two consecutive vehicles following the same route share both the upstream and downstream sections, whereas vehicles entering from different branches share only the downstream section after the merging point. A conservative but inefficient solution is to delay vehicle $i+1$ until vehicle $i$ has completely traversed all corridor sections shared by the two vehicles. If these shared sections are indexed by $j \in \mathcal{J}^{\mathrm{shared}}_{(i,i+1)}$, this corresponds to choosing
% \begin{equation}\label{eq:conservative gap}
%     \Delta t^{m}_{(i,i+1)}
%     >
%     \sum_{j \in \mathcal{J}^{\mathrm{shared}}_{(i,i+1)}} \tau_j .
% \end{equation} 
% This choice eliminates overlap between the two vehicles on the shared sections and is therefore safe, but it can substantially reduce throughput. 

The set of corridor sections shared by two consecutive vehicles depends on their entry CWPs. As shown in Fig.~\ref{fig:UAM_merge}, vehicles with the same entry CWP share both the upstream and downstream sections, while vehicles with different entry CWPs share only the downstream section after the merging point. Let $\mathcal{J}^{\mathrm{shared}}_{(i,i+1)}$ denote the set of corridor-section indices shared by vehicles $i$ and $i+1$.
Then, a conservative but inefficient solution to Problem~\ref{probl:etagap_merge} is to delay vehicle $i+1$ until vehicle $i$ has completely traversed all sections in $\mathcal{J}^{\mathrm{shared}}_{(i,i+1)}$. 
This corresponds to choosing
\begin{equation}\label{eq:conservative gap}
    \Delta t^{m}_{(i,i+1)}
    \ge
    \sum_{j \in \mathcal{J}^{\mathrm{shared}}_{(i,i+1)}} \tau_j .
\end{equation}
This choice of ETA gap avoids overlap on the shared sections and is thus safe, but can substantially reduce traffic throughput.

 Therefore, our goal is instead to compute the smallest ETA gap at the merging point that guarantees~\eqref{eq distance} for every consecutive vehicle pair on each shared corridor section. 
In Sections~\ref{sec:wc_eta_gap} and~\ref{sec:stoc_eta_gap}, we consider two types of trajectory bounds on vehicle trajectories for this purpose, and formulate the corresponding ETA-gap optimization problems in~\eqref{eq:wc_eta_gap_opt} and~\eqref{eq:stoc_eta_gap_opt}. 
These computed sufficient ETA gaps are then used in Section~\ref{sec:entrance_scheduling} to construct a first-come, first-served corridor entrance schedule based on vehicle operators' requested ETAs.

\section{Worst-Case Trajectory Bound and ETA-Gap}
\label{sec:wc_eta_gap}

This section first constructs worst-case spatio-temporal bounds on vehicle trajectories under the speed-limit constraint~\eqref{eq:vlim}. 
These bounds are then used to compute a sufficient ETA gap for Problem~\ref{probl:etagap_merge}. 
The construction in this section follows the same technical idea as those developed for a single corridor in~\cite{pruekprasert2025safe} and sequential corridors in~\cite{pruekprasert2026ifac}, but is tailored here to the merging-corridor setting considered in this paper.

\subsection{Spatio-Temporal Bound Construction}
\label{subsec:wc_bounds} 

% We construct worst-case bounds on shared corridor sections using the speed limits $\vMinj$ and $\vMaxj$ in~\eqref{eq:vlim} and the nominal traversal times $\tau_j$ in~\eqref{eq:tau} for each corridor section $j$ introduced in Section~\ref{subsec:merging}. 
% These bounds enclose admissible vehicle positions and are used for ETA-gap computation.
Consider a corridor section~$j$ of length $l_j$, with speed limits $\vMinj$ and $\vMaxj$ and nominal traversal time $\tau_j$, shared by two consecutive vehicles $i$ and $i+1$. 
Vehicle~$i$ is scheduled to enter section~$j$ at time $t_{i,j}$ and leave it at time $t_{i,j+1}=t_{i,j}+\tau_j$. 
%Throughout the section, 
The vehicles must satisfy the speed-limit constraint~\eqref{eq:vlim}.

Among all such admissible vehicle trajectories, we construct two extreme profiles. 
The first starts at the minimum speed $\vMinj$ and then switches to the maximum speed $\vMaxj$, yielding a lower position bound. 
The second starts at $\vMaxj$ and then switches to $\vMinj$, yielding an upper position bound. 
These profiles are conservative under the simplifying assumption of instantaneous speed changes.
A different worst-case construction that also incorporates the acceleration constraint~\eqref{eq:acc_bounds} was studied in~\cite{pruekprasert2025safe}. 
In this paper, however, we consider only the speed-limit constraints, as in~\cite{pruekprasert2026ifac}, to keep the ETA-gap computation efficient.

% Let $p_j$ denote the position of the entry CWP of section~$j$ along the corresponding route. 
% For the slow-to-fast extreme profile, define the speed-changing time
% \[
%     \widecheck{t}_{j}
%     =
%     \frac{l_j - \vMaxj \tau_j}{\vMinj - \vMaxj},
% \]
% which is well defined as \(l_j/\vMaxj \le \tau_j \le l_j/\vMinj\). 
% The corresponding lower worst-case position bound for vehicle~$i$ on section~$j$ is
% \begin{equation}
% \label{eq:wc_lower_bound}
% \underline{x}^{j,\mathrm{wc}}_{i}(t)
% =
% \begin{cases}
% p_j + \vMinj (t - t_{i,j}),
% &
% \begin{aligned}
% &\text{if }t_{i,j} \le t \\
% &\le t_{i,j} + \widecheck{t}_{j},
% \end{aligned}
% \\ 
% \begin{aligned}
% &p_j + \vMinj \widecheck{t}_{j} \\
% &\quad + \vMaxj (t - t_{i,j} - \widecheck{t}_{j}),
% \end{aligned}
% &
% \begin{aligned}
% &\text{if }t_{i,j} + \widecheck{t}_{j} < t \\
% &\le t_{i,j+1}.
% \end{aligned}
% \end{cases}
% \end{equation}
% The upper worst-case bound $\overline{x}^{j,\mathrm{wc}}_{i}(t)$ is obtained analogously by using the fast-to-slow profile, i.e., by interchanging $\vMinj$ and $\vMaxj$ in~\eqref{eq:wc_lower_bound}. 
% Thus, any admissible trajectory satisfies
% \begin{equation}
% \label{eq:wc_bound_enclosure}
%     \underline{x}^{j,\mathrm{wc}}_{i}(t)
%     \le
%     x_i(t)
%     \le
%     \overline{x}^{j,\mathrm{wc}}_{i}(t),
%     \qquad
%     t \in [t_{i,j}, t_{i,j+1}].
% \end{equation}

\begin{figure}[tbp]
\centering

\begin{subfigure}{0.8\columnwidth}
    \centering
    \includegraphics[width=\linewidth]{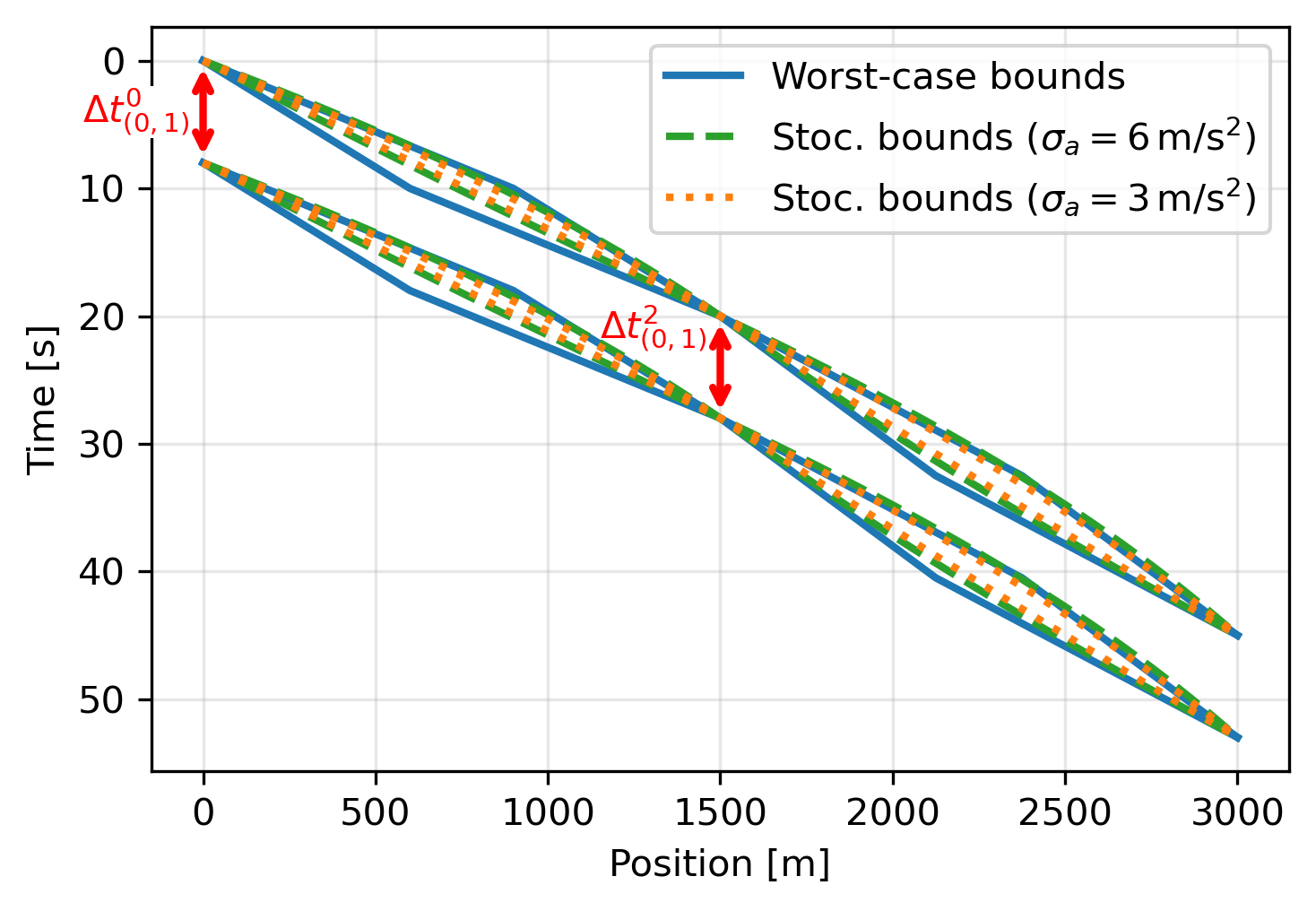}
    \caption{}
    \label{fig:bounds_a}
\end{subfigure}

\begin{subfigure}{0.8\columnwidth}
    \centering
    \includegraphics[width=\linewidth]{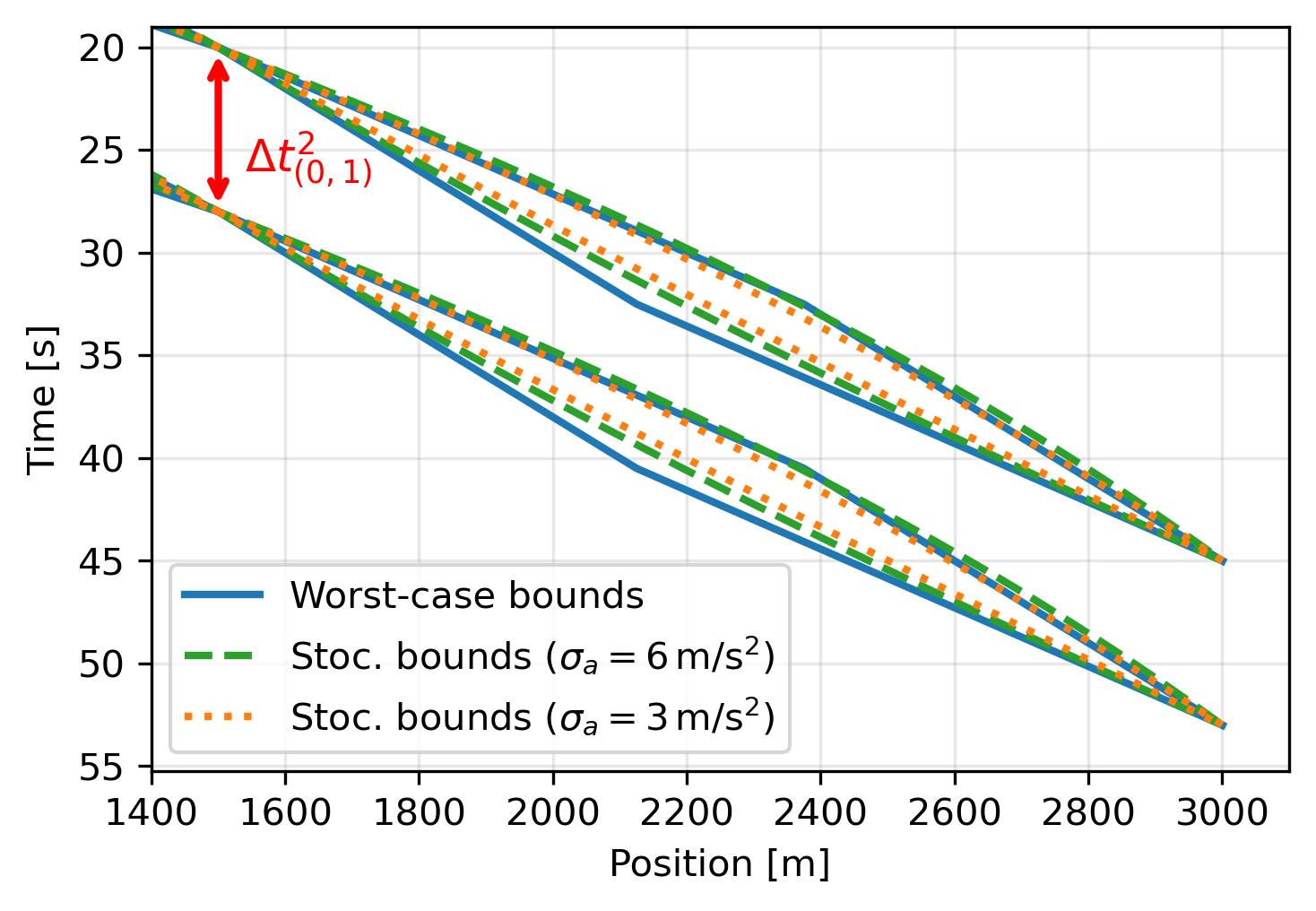}
    \caption{}
    \label{fig:bounds_b}
\end{subfigure}

% \caption{Spatio-temporal trajectory bounds for two vehicles entering the corridor network consecutively from the same CWP. 
% (a)~Worst-case bounds (blue solid lines), stochastic bounds with $\sigma_a=3$\,m/s$^2$ (orange dashed lines), and stochastic bounds with $\sigma_a=6$\,m/s$^2$ (green dotted lines). 
% (b)~Close-up view in the second corridor section.}
\caption{Spatio-temporal trajectory bounds for two vehicles entering consecutively from the same CWP. 
(a)~Worst-case bounds (blue solid lines), stochastic bounds with $\sigma_a=3$\,m/s$^2$ (orange dashed lines), and stochastic bounds with $\sigma_a=6$\,m/s$^2$ (green dotted lines). 
(b)~Close-up of the second section.}
\label{fig:bounds}
\end{figure}

Let $p_j$ denote the position of the entry CWP of section~$j$ along the corresponding route. 
For the slow-to-fast and fast-to-slow extreme profiles, define the speed-changing times
\[
    \widecheck{t}_{j}
    =
    \frac{l_j - \vMaxj \tau_j}{\vMinj - \vMaxj},
    \qquad
    \widehat{t}_{j}
    =
    \frac{l_j - \vMinj \tau_j}{\vMaxj - \vMinj}.
\]
which are well defined since $\tau_j$ is computed from the section-average speed as in \eqref{eq:tau}, and therefore satisfies $l_j/\vMaxj \le \tau_j \le l_j/\vMinj$.
The corresponding lower worst-case position bound for vehicle~$i$ on section~$j$ is
\begin{equation}
\label{eq:wc_lower_bound}
\underline{x}^{j,\mathrm{wc}}_{i}(t)
=
\begin{cases}
p_j + \vMinj (t - t_{i,j}),
&
\begin{aligned}
&\text{if }t_{i,j} \le t \\
&\le t_{i,j} + \widecheck{t}_{j},
\end{aligned}
\\ 
\begin{aligned}
&p_j + \vMinj \widecheck{t}_{j} \\
&\quad + \vMaxj (t - t_{i,j} - \widecheck{t}_{j}),
\end{aligned}
&
\begin{aligned}
&\text{if }t_{i,j} + \widecheck{t}_{j} < t \\
&\le t_{i,j+1}.
\end{aligned}
\end{cases}
\end{equation}
The upper worst-case bound $\overline{x}^{j,\mathrm{wc}}_{i}(t)$ is obtained analogously by switching $\vMinj$ and $\vMaxj$ in~\eqref{eq:wc_lower_bound}, and replacing $\widecheck{t}_{j}$ with $\widehat{t}_{j}$. 
Thus, any admissible trajectory satisfies
\begin{equation}
\label{eq:wc_bound_enclosure}
    \underline{x}^{j,\mathrm{wc}}_{i}(t)
    \le
    x_i(t)
    \le
    \overline{x}^{j,\mathrm{wc}}_{i}(t),
    \qquad
    t \in [t_{i,j}, t_{i,j+1}].
\end{equation}

For multiple corridor sections, the section-wise lower bounds in~\eqref{eq:wc_lower_bound} and their corresponding upper bounds are computed separately and then shifted by the corresponding CWP positions to form route-wise worst-case bounds.

Fig.~\ref{fig:bounds} depicts worst-case bounds for two consecutive vehicles sharing the same entry CWP, shown as blue solid lines.

\subsection{Safe ETA-Gap Computation}
\label{subsec:wc_eta_gap_computation}

We now compute a sufficient ETA gap using the worst-case position bounds. 
The worst-case bounds can be evaluated in continuous time. 
However, the numerical simulation in Section~\ref{sec:simulation} evolves vehicle states in discrete time. 
We therefore include an additional distance margin $d_{\mathrm{margin}}$ to account for execution delay and possible inter-sample loss of separation.

For two consecutive vehicles $i$ and $i+1$, we first compute the section-wise worst-case bounds for each shared corridor section $j\in\mathcal{J}^{\mathrm{shared}}_{(i,i+1)}$. 
These section-wise bounds are then shifted by the corresponding CWP positions and connected to form route-wise worst-case lower and upper bounds over the shared corridor, denoted by $\underline{x}^{\mathrm{wc}}_{i}(t)$ and $\overline{x}^{\mathrm{wc}}_{i+1}(t)$, respectively. 
The sufficient worst-case safety condition is
\begin{equation}
\label{eq:wc_safe_condition}
    \underline{x}^{\mathrm{wc}}_{i}(t)
    -
    \overline{x}^{\mathrm{wc}}_{i+1}(t)
    -
    d_{\mathrm{margin}}
    \ge
    d_{\mathrm{safe}} .
\end{equation}
If~\eqref{eq:wc_safe_condition} holds whenever both vehicles are on the shared route, then it serves as a sufficient condition for the original separation requirement~\eqref{eq distance}, provided that $d_{\mathrm{margin}}$ covers execution delay and possible inter-sample loss of separation.

To compute the gap efficiently, we use the piecewise-affine structure of the worst-case bounds. 
For each shared section~$j$, the slope of $\underline{x}^{j,\mathrm{wc}}_{i}(t)$ changes at $t_{i,j}+\widecheck{t}_{j}$, whereas the slope of $\overline{x}^{j,\mathrm{wc}}_{i+1}(t)$ changes at $t_{i+1,j}+\widehat{t}_{j}$.
Hence, the left-hand side of~\eqref{eq:wc_safe_condition} is piecewise affine, so its minimum occurs at an endpoint or switching time. 
It is therefore sufficient to check~\eqref{eq:wc_safe_condition} only at the set of critical times
\begin{equation}
    \label{eq:critical_times}
\begin{aligned}
\widetilde{\mathcal{T}}_{(i,i+1)}
=
\bigcup_{j\in\mathcal{J}^{\mathrm{shared}}_{(i,i+1)}}
\{&
t_{i,j},\,
t_{i,j}+\widecheck{t}_{j},\,
t_{i,j+1},\\
&
t_{i+1,j},\,
t_{i+1,j}+\widehat{t}_{j},\,
t_{i+1,j+1}
\}.
\end{aligned}
\end{equation}
These are the entry, switching, and exit times of the two vehicles over the shared sections. 
A similar piecewise-affine bounding argument was used for sequential corridors in~\cite{pruekprasert2026ifac}.

The safe ETA-gap computation can then be written as
\begin{align}
\label{eq:wc_eta_gap_opt}
\begin{split}
\underset{\Delta t^{m}_{(i,i+1)}}{\min}\quad
& \Delta t^{m}_{(i,i+1)} \\
\text{subject to}\quad
& \underline{x}^{\mathrm{wc}}_{i}(t)
-
\overline{x}^{\mathrm{wc}}_{i+1}(t)
-
d_{\mathrm{margin}}
\ge d_{\mathrm{safe}},
\\
&\forall\, t \in \widetilde{\mathcal{T}}_{(i,i+1)}, 
\quad
% \Delta t^{m}_{(i,i+1)} = t_{i+1,m}-t_{i,m},\\ 
%
\\&
0 \le \Delta t^{m}_{(i,i+1)}
\le
\sum_{j\in\mathcal{J}^{\mathrm{shared}}_{(i,i+1)}}\tau_j .
\end{split}
\end{align}

Here, $\Delta t^{m}_{(i,i+1)}$ is the ETA gap between the two consecutive vehicles at the merging point $\CWP{m}$.
The first constraint enforces the conservative separation condition~\eqref{eq:wc_safe_condition} over the shared route; therefore, any feasible solution of~\eqref{eq:wc_eta_gap_opt} is a sufficient ETA gap for Problem~\ref{probl:etagap_merge}, provided that $d_{\mathrm{margin}}$ covers execution delay and possible inter-sample loss of separation. 
The upper bound in~\eqref{eq:wc_eta_gap_opt} corresponds to the conservative case~\eqref{eq:conservative gap}, where the following vehicle enters the shared portion only after the leading vehicle has traversed all shared sections.

Since the decision variable is the scalar ETA gap $\Delta t^{m}_{(i,i+1)}$, the optimization problem in~\eqref{eq:wc_eta_gap_opt} can be solved using standard off-the-shelf numerical optimization tools. 
In the simulation study in Section~\ref{sec:simulation}, we use the SciPy function \texttt{scipy.optimize.minimize\_scalar}.

\section{Stochastic Trajectory Bound and ETA-Gap}
\label{sec:stoc_eta_gap}

This section constructs stochastic spatio-temporal bounds on vehicle trajectories under the discretized kinematic model and uses them to compute a sufficient ETA gap for Problem~\ref{probl:etagap_merge}. 
% Unlike the worst-case bounds in Section~\ref{sec:wc_eta_gap}, the bounds here are derived from a stochastic acceleration model and provide high-probability position envelopes.

\subsection{Stochastic Spatio-Temporal Bound Construction}
\label{subsec:stoc_bounds}

We use a discrete-time version of the kinematic model in~\eqref{eq:kinematic}. 
Consider corridor section~$j$ with length $l_j$ and traversal time $\tau_j$. 
We assume known nominal entry and exit speeds $v^0_j$ and $v^E_j$, both with standard deviation $\sigma_v$. 
For simplicity, we consider position relative to the entry point of the corridor section, i.e., the section starts at position $0$.

Let $\delta_t$ be the time discretization step, and let $N_j=\mathrm{round}(\tau_j/\delta_t)$ so that $N_j\delta_t \approx \tau_j$. 
At time step $k \in \{0,\dots,N_j\}$, the state of vehicle~$i$ is
$z_{i,k} = [x_{i,k},\, v_{i,k}]^\top$,
where $x_{i,k}$ and $v_{i,k}$ denote the along-section position and speed, respectively.
The discretized dynamics are
\begin{equation}
\label{eq:stoc_discrete_dynamics}
\begin{aligned}
    z_{i,k+1}
    &=
    A z_{i,k}
    +
    B(a^{\mathrm{nom}}_j + w_k)\\
    &=A z_{i,k}
    +
    B a^{\mathrm{nom}}_j + B w_k,
\end{aligned} 
\end{equation}
where
$A =
\begin{bmatrix}
    1 & \delta_t \\
    0 & 1
\end{bmatrix}$
and
$B =
\begin{bmatrix}
    \frac{1}{2}\delta_t^2 \\
    \delta_t
\end{bmatrix}$.
Here, $a^{\mathrm{nom}}_j$ is the nominal constant acceleration over section~$j$, %connecting the prescribed entry and exit speeds $v^0_j$ and $v^E_j$,
given by
\begin{equation}
\label{eq:a_nom}
    a^{\mathrm{nom}}_j
    =
    \frac{v^E_j - v^0_j}{N_j\delta_t}.
\end{equation}

The noise $w_k$ models uncertainty in acceleration. 
It is treated as a zero-mean Gaussian disturbance with variance $\sigma_a^2$, i.e., $w_k \sim \mathcal{N}(0,\sigma_a^2)$. 
The induced discrete-time process-noise covariance for the noise term $B w_k$ is
\begin{equation*}
% \label{eq:stoc_Q}
    Q
    =
    \sigma_a^2 B B^\top
    =
    \sigma_a^2
    \begin{bmatrix}
        \delta_t^4/4 & \delta_t^3/2 \\
        \delta_t^3/2 & \delta_t^2
    \end{bmatrix}.
\end{equation*}

% Thus, $Q$ captures how uncertainty in acceleration affects the covariance of the position and speed states over one time step. 
% In simulation, the realized acceleration is further truncated to satisfy~\eqref{eq:acc_bounds}, but the stochastic-bound construction itself uses the Gaussian covariance model above.

We compute the stochastic trajectory bounds in four steps: forward propagation, terminal conditioning, backward smoothing, and pointwise bound construction, following standard filtering and smoothing ideas~\cite{kalman1960new,rauch1965maximum,sarkka2013bayesian}.

First, starting from the entry CWP of section~$j$, we propagate the state mean and covariance forward in time. 
Let $\mu_{i,k}$ and $P_{i,k}$ denote the mean and covariance of $z_{i,k}$, respectively. 
Since the initial position is $0$ and the nominal entry speed is $v^0_j$ with variance $\sigma_v^2$, the initial mean is set to the nominal entry state,
$\mu_{i,0}
=
\begin{bmatrix}
    0 &
    v^0_j
\end{bmatrix}^\top$,
and uncertainty in the entry speed is incorporated through the covariance matrix
$P_{i,0}
=
\begin{bmatrix}
    0 & 0 \\
    0 & \sigma_v^2
\end{bmatrix}$. %,
%where $\sigma_v$ is the standard deviation used for the entry-speed uncertainty. 
Then, for $k=0,\dots,N_j-1$, the forward propagation is
\[\mu_{i,k+1}
=
A\mu_{i,k} + B a^{\mathrm{nom}}_j, \quad P_{i,k+1}
=
A P_{i,k} A^\top + Q\]
Hence, the covariance matrix $P_{i,k}$ is updated at each step by adding the process-noise covariance $Q$, so that acceleration uncertainty accumulates along the section.

Second, we consider the terminal state at $k=N_j$ obtained from the forward propagation. 
Since $l_j$ is the end position of the corridor section and $v^E_j$ is the nominal exit speed with variance $\sigma_v^2$, we define
$
y_j =
\begin{bmatrix}
l_j \\
v^E_j
\end{bmatrix}
$
and
$
R_j =
\begin{bmatrix}
\varepsilon & 0 \\
0 & \sigma_v^2
\end{bmatrix},
$
where $\varepsilon$ is a small constant for terminal-position imprecision.
Treating $y_j$ as a terminal observation with covariance matrix $R_j$, and using the observation matrix $H=I$ (the $2\times 2$ identity matrix), we define the innovation covariance and Kalman gain by the standard linear-Gaussian update~\cite{kalman1960new,sarkka2013bayesian} as follows:
\[
S_j = H P_{i,N_j} H^\top + R_j,
\qquad
K_j = P_{i,N_j} H^\top S_j^{-1}.
\]
Then, using the standard linear-Gaussian conditioning formulas, the terminal mean and covariance are updated as
\[
    \mu_{i,N_j}
    \gets
    \mu_{i,N_j} + K_j\bigl(y_j - \mu_{i,N_j}\bigr),
    \qquad
    P_{i,N_j}
    \gets
    (I-K_j)P_{i,N_j}.
\]
This update incorporates the terminal information that the position is near $l_j$ and the speed is near $v^E_j$.

Third, we apply Rauch--Tung--Striebel smoothing~\cite{rauch1965maximum,sarkka2013bayesian} using the smoothing gain
\[
G_{i,k}
=
P_{i,k} A^\top \bigl(P_{i,k+1}\bigr)^{-1},
\qquad k=N_j-1,\dots,0.
\]
The smoothed mean $\mu^{\mathrm{sm}}_{i,k}$ and covariance $P^{\mathrm{sm}}_{i,k}$ are then propagated backward from $k=N_j$ to $k=0$ as
\[
    \mu^{\mathrm{sm}}_{i,k}
    =
    \mu_{i,k}
    +
    G_{i,k}
    \Bigl(
        \mu^{\mathrm{sm}}_{i,k+1}
        -
        \bigl(A\mu_{i,k}+Ba^{\mathrm{nom}}_j\bigr)
    \Bigr),
\]
\[
    P^{\mathrm{sm}}_{i,k}
    =
    P_{i,k}
    +
    G_{i,k}
    \bigl(
        P^{\mathrm{sm}}_{i,k+1}
        -
        P_{i,k+1}
    \bigr)
    G_{i,k}^\top.
\]
Here, the smoothed covariance $P^{\mathrm{sm}}_{i,k}$ directly determines the width of the stochastic position bounds at time step~$k$.

Finally, for a desired probability level $\rho \in (0,1)$, we construct whole-trajectory stochastic position bounds over the $N_j+1$ time steps.
Let $x^{\mathrm{sm}}_{i,k}$ denote the position component of $\mu^{\mathrm{sm}}_{i,k}$, and let $\sigma_{x,k}=\sqrt{(P^{\mathrm{sm}}_{i,k})_{11}}$ be its standard deviation.
Accordingly, we want the position to remain within the stochastic bounds at every time step with probability at least $\rho$, i.e.,
\[
\Pr\!\left(
\bigcap_{k=0}^{N_j}
\left\{
\lvert x_{i,k}-x^{\mathrm{sm}}_{i,k}\rvert \le z_k
\right\}
\right)
\ge \rho.
\]
Let $E_k=\{\lvert x_{i,k}-x^{\mathrm{sm}}_{i,k}\rvert > z_k\}$ be the violation event at time step $k$. 
Then it is sufficient to require
\[
\Pr\!\left(
\bigcup_{k=0}^{N_j} E_k
\right)
\le 1-\rho.
\]
By the union bound (Boole's inequality),
\begin{equation}\label{eq:union bound}
    \Pr\!\left(
\bigcup_{k=0}^{N_j} E_k
\right)
\le
\sum_{k=0}^{N_j}\Pr(E_k).
\end{equation}
Thus, it is enough to impose the same one-step failure bound $\Pr(E_k)\le \alpha$ for all $k$, where $\alpha=\frac{1-\rho}{N_j+1}$, so that the union bound in~\eqref{eq:union bound} gives
$\Pr\!\left(\bigcup_{k=0}^{N_j} E_k\right)\le (N_j+1)\alpha = 1-\rho$.
% Thus, it is enough to impose the same one-step failure bound $\Pr(E_k)\le \alpha$ for all $k$,
% where $\alpha=\frac{1-\rho}{N_j+1}$, since $(N_j+1)\alpha \le 1-\rho$.
% \[
% \alpha=\frac{1-\rho}{N_j+1} \quad \text{ as } \quad
% (N_j+1)\alpha \le 1-\rho.
% \]

% Since $x_{i,k}$ is modeled as Gaussian with mean $x^{\mathrm{sm}}_{i,k}$ and standard deviation $\sigma_{x,k}$, we have $\frac{x_{i,k}-x^{\mathrm{sm}}_{i,k}}{\sigma_{x,k}} \sim \mathcal{N}(0,1)$ and
% $\Pr\!\left(
% \left| \frac{x_{i,k}-x^{\mathrm{sm}}_{i,k}}{\sigma_{x,k}}\right|
% \leq \Phi^{-1}\!\left(1-\frac{\alpha}{2}\right)
% \right)= 1-\alpha$, where $\Phi^{-1}$ is the inverse standard normal cumulative distribution function. 

% We now choose $z_k$ so that each one-step violation probability is at most $\alpha$. 
% Since $x_{i,k}$ is modeled as Gaussian with mean $x^{\mathrm{sm}}_{i,k}$ and standard deviation $\sigma_{x,k}$, choosing
% $z_k = \Phi^{-1}\!\left(1-\frac{\alpha}{2}\right)\sigma_{x,k}$
% gives the two-sided Gaussian bound
% \[
% \Pr\!\left(
% \lvert x_{i,k}-x^{\mathrm{sm}}_{i,k}\rvert \le z_k
% \right)\ge 1-\alpha,
% \]
% which is equivalent to the desired condition $\Pr(E_k)\le \alpha$.

Since $x_{i,k}$ is modeled as Gaussian with mean $x^{\mathrm{sm}}_{i,k}$ and standard deviation $\sigma_{x,k}$, we have
$\frac{x_{i,k}-x^{\mathrm{sm}}_{i,k}}{\sigma_{x,k}} \sim \mathcal{N}(0,1)$.
We therefore choose $z_k$ so that each one-step violation probability is at most $\alpha$. 
Specifically, letting
\[z_k = \Phi^{-1}\!\left(1-\frac{\alpha}{2}\right)\sigma_{x,k},\]
where $\Phi^{-1}$ is the inverse standard normal cumulative distribution function, gives
\[
\Pr\!\left(
\lvert x_{i,k}-x^{\mathrm{sm}}_{i,k}\rvert \le z_k
\right)= 1-\alpha,
\]
which implies $\Pr(E_k)=\alpha$, and therefore $\Pr(E_k)\le \alpha$.

Accordingly, the stochastic bounds are
\begin{equation}
\label{eq:stoc_position_tube}
    \underline{x}^{j,\mathrm{stoc}}_{i,k}
    =
    x^{\mathrm{sm}}_{i,k} - z_k,
    \qquad
    \overline{x}^{j,\mathrm{stoc}}_{i,k}
    =
    x^{\mathrm{sm}}_{i,k} + z_k,
\end{equation}
The trajectory remains within these bounds with probability at least $\rho$ under the assumed stochastic model.

For multiple corridor sections, the section-wise bounds in~\eqref{eq:stoc_position_tube} are computed separately and then shifted by the corresponding CWP positions to form route-wise stochastic bounds. 
Fig.~\ref{fig:bounds} also illustrates stochastic bounds for two acceleration uncertainty levels: $\sigma_a=3$\,m/s$^2$, shown by orange dashed lines, and $\sigma_a=6$\,m/s$^2$, shown by green dotted lines.
As expected, the bounds widen as the assumed acceleration uncertainty increases. 

\subsection{Stochastic ETA-Gap Computation}
\label{subsec:stoc_eta_gap_computation}

We now compute a sufficient ETA gap using the stochastic position bounds. 
As in the worst-case bounds computation, we include the additional distance margin $d_{\mathrm{margin}}$ for execution delay and grid-based safety checking. 
For two consecutive vehicles $i$ and $i+1$, we first compute the section-wise stochastic bounds in~\eqref{eq:stoc_position_tube} for each shared corridor section $j\in\mathcal{J}^{\mathrm{shared}}_{(i,i+1)}$. 
These section-wise bounds are then shifted by the corresponding CWP positions and concatenated to form route-wise stochastic lower and upper bounds over the shared corridor, denoted by $\underline{x}^{\mathrm{stoc}}_{i,k}$ and $\overline{x}^{\mathrm{stoc}}_{i+1,k}$, respectively.

For a candidate merging ETA gap, the follower bound is shifted in time by this gap, and the stochastic separation condition is checked over all overlapping time steps. 
The sufficient stochastic safety condition is
\begin{equation}
\label{eq:stoc_safe_condition}
    \underline{x}^{\mathrm{stoc}}_{i,k}
    -
    \overline{x}^{\mathrm{stoc}}_{i+1,k}
    -
    d_{\mathrm{margin}}
    \ge
    d_{\mathrm{safe}} .
\end{equation}
% Equivalently, the stochastic tube must provide a separation of at least $d_{\mathrm{safe}} + d_{\mathrm{margin}}$ at every checked time steps.

The resulting ETA-gap computation is
\begin{align}
\label{eq:stoc_eta_gap_opt}
\begin{split}
\underset{\Delta t^{m,\mathrm{stoc}}_{(i,i+1)}}{\min}\quad
& \Delta t^{m,\mathrm{stoc}}_{(i,i+1)} \\
\text{subject to}\quad
& \underline{x}^{\mathrm{stoc}}_{i,k}
-
\overline{x}^{\mathrm{stoc}}_{i+1,k}
-
d_{\mathrm{margin}}
\ge d_{\mathrm{safe}},
\quad \forall\, k, 
% \\
% & \Delta t^{m,\mathrm{stoc}}_{(i,i+1)}
% =
% t_{i+1,m}-t_{i,m},
\\
& 0 \le \Delta t^{m,\mathrm{stoc}}_{(i,i+1)}
\le
\sum_{j\in\mathcal{J}^{\mathrm{shared}}_{(i,i+1)}}\tau_j .
\end{split}
\end{align}
Here, $k$ ranges over all time steps where the two shifted route-wise stochastic bounds overlap on the shared corridor. 
Thus, the first constraint enforces separation not only when both vehicles are on the same shared section, but also when they occupy different shared sections of the common route. 
Accordingly, any feasible solution of~\eqref{eq:stoc_eta_gap_opt} gives a stochastic sufficient ETA gap for Problem~\ref{probl:etagap_merge}, provided that $d_{\mathrm{margin}}$ covers execution delay and inter-sample loss of separation, and that the assumed stochastic model is appropriate.

Since the decision variable is the scalar ETA gap, the search can be implemented using standard one-dimensional numerical methods. 
In the simulation study in Section~\ref{sec:simulation}, we use a bisection search over the ETA gap. 
The initial search interval is set to
\([0,\,\sum_{j \in \mathcal{J}^{\mathrm{shared}}_{(i,i+1)}}\tau_j]\), namely between zero and the conservative upper bound in~\eqref{eq:conservative gap}. 
At each iteration, we test the midpoint $\frac{\mathit{inv}_\mathrm{lb}+\mathit{inv}_\mathrm{ub}}{2}$ of the current interval $[\mathit{inv}_\mathrm{lb}, \mathit{inv}_\mathrm{ub}]$. 
If the stochastic separation condition is satisfied, then the midpoint is feasible, and we update the search interval to
$[\mathit{inv}_\mathrm{lb}, \frac{\mathit{inv}_\mathrm{lb}+\mathit{inv}_\mathrm{ub}}{2}]$
in order to search for a smaller feasible gap. 
Otherwise, the midpoint is infeasible, and we update the interval to
$[\frac{\mathit{inv}_\mathrm{lb}+\mathit{inv}_\mathrm{ub}}{2}, \mathit{inv}_\mathrm{ub}]$. 
This process is repeated until the interval width is below a prescribed tolerance.

% Compared with the worst-case speed-profile bound, the stochastic bound is generally less conservative when the acceleration uncertainty is moderate. 
% However, its guarantee depends on the assumed stochastic model, including the values of $\sigma_a$ and $\sigma_v$, the terminal conditioning and smoothing procedure, the probability level $\rho$, and the additional margin $d_{\mathrm{margin}}$.
Compared with the worst-case speed-profile bound, the stochastic bound is generally less conservative when the acceleration uncertainty is moderate. 
However, its guarantee depends on the assumed stochastic model, the probability level $\rho$, and the additional margin $d_{\mathrm{margin}}$.

\section{Corridor Entrance Scheduling}
\label{sec:entrance_scheduling}

We now describe how the safe ETA gaps computed in Sections~\ref{sec:wc_eta_gap} and~\ref{sec:stoc_eta_gap} are used for corridor ETA scheduling.
The scheduling is performed in a first-come, first-served order at the merging point $\CWP{m}$. 
For each vehicle, the PSU checks the required ETA gap with the immediately preceding vehicle at the merge and, if applicable, with the most recent previously scheduled vehicle that uses the same entry CWP. 

Let $\beta \in \{\mathrm{wc},\mathrm{stoc}\}$ denote the selected bound type, and let
$\Gamma^{\beta}_{a,b}\bigl(\mathcal{J}^{\mathrm{shared}}_{(a,b)}\bigr)$
be the sufficient ETA gap obtained by solving the corresponding ETA-gap optimization problem for vehicles $a$ and $b$ over their shared section set $\mathcal{J}^{\mathrm{shared}}_{(a,b)}$. 
Specifically, $\Gamma^{\mathrm{wc}}_{a,b}(\cdot)$ is obtained from~\eqref{eq:wc_eta_gap_opt}, while $\Gamma^{\mathrm{stoc}}_{a,b}(\cdot)$ is obtained from~\eqref{eq:stoc_eta_gap_opt}. 
The ETA gap is computed using the spatio-temporal bounds over all shared corridor sections. 
If two vehicles use the same entry CWP, then the shared section set includes both the upstream and downstream corridor sections. 
If they enter from different branches, then it includes only the downstream section after the merging point.

% Once the scheduled ETA at the $\CWP{m}$ is determined, the scheduled entrance ETA can be obtained by shifting backward along the vehicle's route. 
% Recall that $c_i$ denotes the entry CWP of vehicle~$i$, and let $\mathcal{J}^{c_i \to m}_i$ be the set of corridor sections from $\CWP{c_i}$ to $\CWP{m}$ on the route of vehicle~$i$. 
% Then
% \[
%     t_{i,c_i}
%     =
%     t_{i,m}
%     -
%     \sum_{j\in\mathcal{J}^{c_i \to m}_i}\tau_j .
% \]
Once the scheduled ETA at $\CWP{m}$ is determined, the entrance ETA is obtained by shifting backward along the incoming branch. 
Recall that $c_i$ denotes the entry CWP of vehicle~$i$, and let $\tau(c_i)$ denote the nominal travel time from $c_i$ to the merging point $\CWP{m}$. 
If the scheduled ETA at $\CWP{m}$ for vehicle~$i$ is $t_{i,m}$, then its ETA at $c_i$ is
$t_{i,c_i}
=
t_{i,m}
-
\tau(c_i)$.

\subsection{Scheduled ETA Computation}
\label{subsec:initial_entrance_schedule}
\begin{algorithm}[tbp]
\caption{First-come, first-served corridor entry scheduling}
\label{alg:entrance_scheduling} 
\begin{algorithmic}[1]
\REQUIRE
\begin{tabular}[t]{@{}l@{}}
Vehicles $i \in \{0,1,\ldots,n-1\}$ ordered by\\\quad first requested arrival at $\CWP{m}$;\\
Entry CWP $c_i$ for each vehicle $i$;\\
Proposed merge ETA $t^{\mathrm{proposed}}_{i,m}$ for each vehicle $i$;\\
Trajectory bound type $\beta\in\{\mathrm{wc},\mathrm{stoc}\}$.
\end{tabular}
\ENSURE Scheduled ETAs at $\CWP{m}$ and entry CWPs

\STATE Set $t_{0,m} \gets t^{\mathrm{proposed}}_{0,m}$.
\STATE Set $t_{0,c_0} \gets t_{0,m} - \tau(c_0)$.
\STATE Set $\mathrm{last}(c_0) \gets 0$.

\FOR{$i=0,\dots,n-2$}
    \STATE Determine $\mathcal{J}^{\mathrm{shared}}_{(i,i+1)}$ from $c_i$ and $c_{i+1}$.
    \STATE Compute $\Delta t_{\mathrm{merge}} \gets \Gamma^{\beta}_{i,i+1}\bigl(\mathcal{J}^{\mathrm{shared}}_{(i,i+1)}\bigr)$.
    \STATE Set $t^{\mathrm{candidate}}_{i+1,m} \gets \max\{t^{\mathrm{proposed}}_{i+1,m},\, t_{i,m}+\Delta t_{\mathrm{merge}}\}$. \label{algline:merge_check}

    \IF{$\mathrm{last}(c_{i+1})$ exists and $\mathrm{last}(c_{i+1})\neq i$}
        \STATE Set $a \gets \mathrm{last}(c_{i+1})$.
        \STATE Determine $\mathcal{J}^{\mathrm{shared}}_{(a,i+1)}$ from $c_a$ and $c_{i+1}$.
        \STATE Compute $\Delta t_{\mathrm{same}} \gets \Gamma^{\beta}_{a,i+1}\bigl(\mathcal{J}^{\mathrm{shared}}_{(a,i+1)}\bigr)$.
        \STATE Set $t^{\mathrm{candidate}}_{i+1,m} \gets \max\{t^{\mathrm{candidate}}_{i+1,m},\, t_{a,m}+\Delta t_{\mathrm{same}}\}$. \label{algline:same_route_check}
    \ENDIF

    \STATE Set $t_{i+1,m} \gets t^{\mathrm{candidate}}_{i+1,m}$.
    \STATE Set $t_{i+1,c_{i+1}} \gets t_{i+1,m} - \tau(c_{i+1})$. \label{algline:backward_shift}
    \STATE Set $\mathrm{last}(c_{i+1}) \gets i+1$. \label{algline:update_last}
\ENDFOR
\end{algorithmic}
\end{algorithm}

We present Algorithm~\ref{alg:entrance_scheduling} for first-come, first-served corridor entry scheduling. 
Here, $t^{\mathrm{proposed}}_{i+1,m}$ denotes the ETA at the merging point $\CWP{m}$ proposed by the operator before adjustment by the PSU. 
Lines~1--2 initialize the schedule for the first vehicle, vehicle $0$, by setting its ETA at the merging point $\CWP{m}$ and computing the corresponding ETA at the entry CWP $c_0$. 
Line~3 stores this vehicle as the most recently scheduled vehicle entering from $c_0$.

The loop in Lines~4--13 then considers the remaining vehicles in order of requested arrival at $\CWP{m}$. 
For vehicle $i+1$, Lines~5--6 compute the ETA gap required with respect to the immediately preceding vehicle in the merge order, based on the corridor sections shared by the pair $(i,i+1)$. 
Line~\ref{algline:merge_check} then sets a tentative ETA at $\CWP{m}$. 

If a previous vehicle with the same entry CWP exists and is not the immediately preceding vehicle in the merge order, Lines~8--11 compute the ETA gap required on the upstream section shared before the merging point. 
Line~\ref{algline:same_route_check} updates the tentative ETA at $\CWP{m}$ if the ETA computed from this new gap is larger than the tentative one previously computed in Line~\ref{algline:merge_check}. 
After the ETA at $\CWP{m}$ is fixed, Line~\ref{algline:backward_shift} computes the corresponding entry-CWP ETA by shifting backward along the incoming branch. 
Finally, Line~\ref{algline:update_last} updates the most recently scheduled vehicle entering from $c_{i+1}$ for later vehicles with the same entry CWP.

\subsection{Schedule Update under Entrance Delay}
\label{subsec:schedule_update}

The schedule above assumes that each vehicle can enter the corridor at its scheduled entrance ETA. 
In execution, however, vehicle~$i$ may be unable to enter at $t_{i,c_i}$.  
Let $\tilde{t}_{i,c_i}$ denote the actual time at which vehicle~$i$ can enter. 
If $\tilde{t}_{i,c_i} > t_{i,c_i}$, we update the entrance ETA to $t_{i,c_i}\gets \tilde{t}_{i,c_i}$. 
Then, the ETA at the merging point is updated by
$ t_{i,m}
    \gets
    t_{i,c_i}
    +
    \tau(c_i)$.
The same entrance delay can also be propagated to the scheduled ETAs of succeeding vehicles so that the updated schedule remains consistent. 
This update also helps preserve scheduling-level safety by maintaining a buffer for subsequent vehicles when disturbances delay corridor entry and shift the realized ETAs.

\section{Numerical Simulation Results}
\label{sec:simulation}

This section evaluates the proposed %ETA coordination 
framework through numerical simulations of the merging corridor scenario. 
% We compare the ETA gaps obtained from the worst-case and stochastic trajectory bounds and examine their effect on corridor scheduling.

\subsection{Simulation Setting}
\label{subsec:simulation_setting}

\subsubsection{Corridor configuration}

We consider the corridor network depicted in Fig.~\ref{fig:UAM_merge}, where traffic from corridor Sections~0 and~1 merges into Section~2 at $\CWP{2}$. 
Each corridor section has length $1{,}500$\,m. 
The speed limits $(\vMinj,\vMaxj)$ in~\eqref{eq:vlim} for Sections~$j=0,1,2$ are $(60,90)$, $(60,80)$, and $(50,70)$\,m/s, respectively. 
Thus, two upstream sections merge into a downstream section with lower admissible speeds, making this scenario particularly challenging for maintaining safe separation and throughput. The acceleration bounds in~\eqref{eq:acc_bounds} are $\bEmergency=-4$\,m/s$^2$ and $\aEmergency=3$\,m/s$^2$. 

We use the discretized dynamics in~\eqref{eq:stoc_discrete_dynamics} with discretization step $\delta_t=0.1$\,s. 
The nominal traversal times, computed from the section lengths and average speeds as in~\eqref{eq:tau}, are rounded to the discretization step, giving
$\tau_0=20.0$\,s, $\tau_1=21.4$\,s, and $\tau_2=25.0$\,s.
For the stochastic bounds, the nominal speeds at the entry and exit CWPs are
$(v^0_0,v^0_1,v^0_2) = (85,75,65)$\,m/s
and
$(v^E_0,v^E_1,v^E_2) = (65,65,55)$\,m/s, respectively, with $\sigma_v = 5$\,m/s, $\varepsilon=10^{-6}$\,m$^2$, and probability level $\rho=0.9$.
For both~\eqref{eq:wc_eta_gap_opt} and~\eqref{eq:stoc_eta_gap_opt}, we set $d_{\mathrm{margin}}=8$\,m.

\subsubsection{Traffic simulation and operational modes}

We simulate a stream of vehicles entering the corridor network over a 600\,s period. 
The entry branch alternates between $\CWP{0}$ and $\CWP{1}$: vehicles with even indices enter from $\CWP{0}$, while vehicles with odd indices enter from $\CWP{1}$. 
All vehicles request the earliest possible ETA at the merging point; equivalently, we set $t^{\mathrm{proposed}}_{i,m}=0$ for all vehicles~$i$, so that the final schedule is determined by the safety-gap constraints.

% We compare four operational modes. 
% The first mode uses the worst-case trajectory bounds from Section~\ref{sec:wc_eta_gap}. 
% The second and third modes use the stochastic trajectory bounds from Section~\ref{sec:stoc_eta_gap} with $\sigma_a=6$\,m/s$^2$ and $\sigma_a=3$\,m/s$^2$, respectively. 
% % For both stochastic-bound modes, the whole-trajectory probability level used to construct the position bounds in~\eqref{eq:stoc_position_tube} is set to $\rho=0.9$.
% The fourth mode is a baseline without ETA coordination at CWPs.

We compare four operational modes. 
The first uses the worst-case trajectory bounds from Section~\ref{sec:wc_eta_gap}. 
The second and third use the stochastic trajectory bounds from Section~\ref{sec:stoc_eta_gap} with $\sigma_a=6$\,m/s$^2$ and $\sigma_a=3$\,m/s$^2$, respectively. 
The fourth is a baseline without ETA coordination at CWPs.

\subsubsection{Vehicle execution model}

For all modes, a vehicle is allowed to enter its entry CWP only if its distance from the preceding vehicle on the same corridor section is greater than $d_{\mathrm{safe}}+d_{\mathrm{margin}}$. 
In the ETA-coordinated modes, if a vehicle cannot enter at its scheduled entrance time because this distance condition is not satisfied, the vehicle keeps its scheduled entry order, but its ETA schedule is updated to account for the resulting delay, as described in Section~\ref{subsec:schedule_update}. 
Such delays can occur when disturbances slow down preceding vehicles, causing deviations from their scheduled ETAs.

Vehicles enter their entry CWPs with speed uncertainty $\sigma_v=5$\,m/s.  
At each simulation time step, the acceleration of each vehicle in corridor section~$j$ is sampled from a Gaussian distribution truncated to $[\bEmergency,\aEmergency]$, with mean $a^{\mathrm{nom}}_j$ as defined in~\eqref{eq:a_nom} and standard deviation $\sigma_a=6$\,m/s$^2$.
For Sections~$j=0,1,2$, the nominal accelerations are
\[
    (a^{\mathrm{nom}}_0,a^{\mathrm{nom}}_1,a^{\mathrm{nom}}_2)
    =
    (-1.00,-0.47,-0.40)\,\mathrm{m/s^2}.
\]
% Fig.~\ref{fig:sigma_a} illustrates the truncated Gaussian acceleration distributions for Section~0 %with $a^{\mathrm{nom}}_0=-1.00$\,m/s$^2$ and
% two different values of $\sigma_a$.

If the distance to the preceding vehicle becomes less than or equal to $d_{\mathrm{safe}}+d_{\mathrm{margin}}$, the vehicle applies braking acceleration $\bEmergency$. 
In the ETA-coordinated modes, vehicles may additionally adjust their acceleration within the limits in~\eqref{eq:acc_bounds} in an attempt to track their scheduled ETAs at CWPs.

% \begin{figure}[tbp]
% \centering

% \begin{subfigure}{0.7\columnwidth}
%     \centering
%     \includegraphics[width=\linewidth]{gaussian_std_for_a2.png}
%     \caption{}
%     \label{fig:sigma_a_a}
% \end{subfigure}

% \begin{subfigure}{0.7\columnwidth}
%     \centering
%     \includegraphics[width=\linewidth]{gaussian_std_for_a.png}
%     \caption{}
%     \label{fig:sigma_a_b}
% \end{subfigure}

% \caption{Gaussian acceleration distributions used in the simulation for corridor Section~0, with mean $a^{\mathrm{nom}}_0=-1$\,m/s$^2$ and acceleration bounds $[\bEmergency,\aEmergency]=[-4,3]$\,m/s$^2$. 
% (a) Standard deviation $\sigma_a=6$\,m/s$^2$. 
% (b) Standard deviation $\sigma_a=3$\,m/s$^2$.}
% \label{fig:sigma_a}
% \end{figure}

% \begin{figure}[tbp]
% \centering
 
%     \centering
%     \includegraphics[width=0.8\linewidth]{gaussian_std_for_a_both.png} 

% \caption{Gaussian acceleration distributions used in the simulation for corridor Section~0, with mean $a^{\mathrm{nom}}_0=-1$\,m/s$^2$ and acceleration bounds $[\bEmergency,\aEmergency]=[-4,3]$\,m/s$^2$. 
% Distributions with standard deviations $\sigma_a=6$\,m/s$^2$ and $\sigma_a=3$\,m/s$^2$ are shown as green dashed and orange dotted lines, respectively.}
% \label{fig:sigma_a}
% \end{figure}

\subsubsection{Disturbance model}

To evaluate robustness under unexpected speed reductions, we introduce disturbances that force vehicles to brake in designated corridor regions. 
These disturbances represent localized events that temporarily require strong deceleration, such as wind effects, operational constraints, or other hazards. 
Specifically, disturbances are applied in the position intervals $[700, 1400]$\,m and $[2200, 2900]$\,m. 
When a disturbance is active, any vehicle located in either interval applies the braking acceleration $\bEmergency$.

We consider six disturbance levels: Level~0 has no disturbance, and each Level~$\ell \in \{1,\dots,5\}$ applies a repeating disturbance pattern with $\ell$\,s active disturbance followed by $(10-\ell)$\,s without disturbance.
These disturbances may cause vehicles to violate the speed-limit assumption in~\eqref{eq:vlim} and deviate from their scheduled ETAs at CWPs. 
They therefore represent off-nominal situations not explicitly accounted for in the original ETA-gap computation. 
When such deviations delay vehicles in the ETA-coordinated modes, the schedule-update rule in Section~\ref{subsec:schedule_update} is applied while preserving the scheduled order. 
Although the precomputed bounds may no longer cover the executed trajectories once the speed-limit assumption is violated, the computed ETA gaps still provide a scheduling-level buffer for updating the plan under disturbances. 
We use these cases to evaluate the robustness of each operational mode when vehicles are forced away from their nominal ETA schedule. 

%\subsubsection{Implementation}
Simulations are implemented in Python~3.12 and run on a MacBook Pro (Apple M4 Max, 64\,GB). 

\subsection{Computed ETA Gaps}
\label{subsec:computed_eta_gaps}

\begin{figure}[tbp]
\centering
\includegraphics[width=0.8\columnwidth]{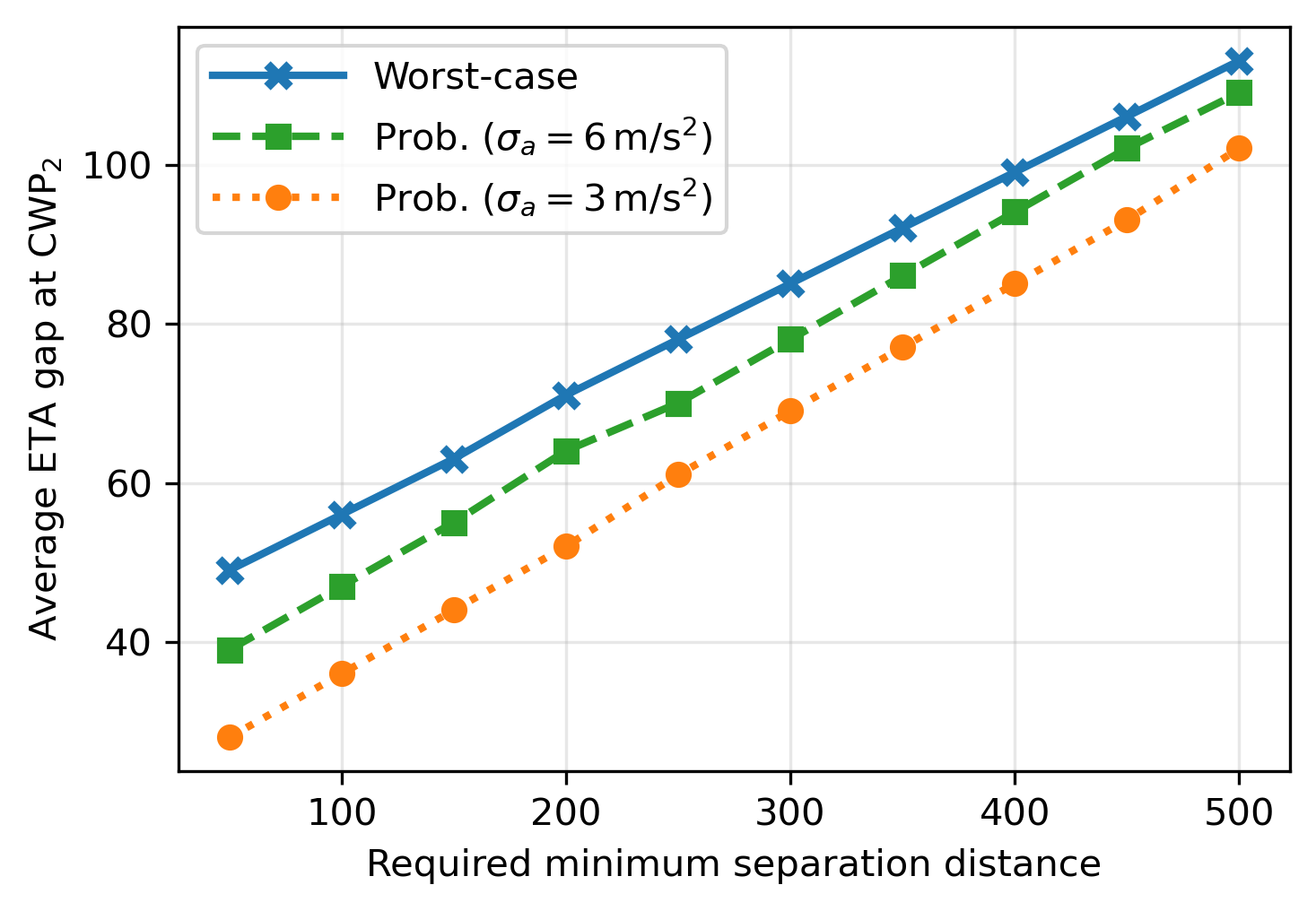}
% \caption{Computed minimum ETA gaps at $\CWP{2}$ for different required separation distances $d_{\mathrm{safe}}$. 
% The ETA gap computed using the worst-case bound is shown by the blue solid line, while the gaps computed using the stochastic bounds with $\sigma_a=6$\,m/s$^2$ and $\sigma_a=3$\,m/s$^2$ are shown by the green dashed and orange dotted lines, respectively.
% }
\caption{Computed minimum ETA gaps at $\CWP{2}$ for different required separation distances $d_{\mathrm{safe}}$. 
The blue solid line shows the worst-case bound, while the green dashed and orange dotted lines show the stochastic bounds with $\sigma_a=6$\,m/s$^2$ and $\sigma_a=3$\,m/s$^2$, respectively.}
\label{fig:computed_eta_gaps}
\end{figure}

We first compare the ETA gaps computed by the three ETA-coordinated modes: the worst-case bounds, the stochastic bounds with $\sigma_a=3$\,m/s$^2$, and with $\sigma_a=6$\,m/s$^2$. 
For each mode, we solve the corresponding problem, \eqref{eq:wc_eta_gap_opt} or~\eqref{eq:stoc_eta_gap_opt}, for
\[
    d_{\mathrm{safe}} \in \{50,100,150,200,250,300,350,400,450,500\}\,\mathrm{m}.
\]
Fig.~\ref{fig:computed_eta_gaps} shows the computed ETA gaps at the merging point.

For all three ETA-coordinated simulation modes, the computed ETA gap increases almost linearly with $d_{\mathrm{safe}}$.
The worst-case bound yields the largest ETA gaps, followed by the stochastic bound with $\sigma_a=6$\,m/s$^2$ and then the stochastic bound with $\sigma_a=3$\,m/s$^2$. 
This ordering is expected: the worst-case bound accounts for extreme admissible speed profiles and is therefore the most conservative, while a larger acceleration uncertainty in the stochastic bound produces a wider position envelope and hence a larger required ETA gap.

The three ETA-gap computations are lightweight in practice. 
In our implementation, solving~\eqref{eq:wc_eta_gap_opt} takes approximately $0.003$\,s per ETA-gap computation, whereas solving~\eqref{eq:stoc_eta_gap_opt} takes approximately $0.012$\,s. 
For~\eqref{eq:wc_eta_gap_opt}, we use the SciPy function \texttt{scipy.optimize.minimize\_scalar}; for~\eqref{eq:stoc_eta_gap_opt}, we use the bisection search over the ETA gap described in Section~\ref{subsec:stoc_eta_gap_computation}. 
The difference in runtime arises because the worst-case computation only evaluates the separation condition at the finite set of critical times in~\eqref{eq:critical_times}, while the stochastic computation checks~\eqref{eq:stoc_safe_condition} over all overlapping discrete time instants of the shifted position bounds. 
Thus, the stochastic bound provides a less conservative ETA gap at the cost of higher, but still modest, computational time.

\subsection{Safety and Throughput Evaluation}
\label{subsec:safety_throughput}
 
We next evaluate the performance of the four simulation modes under the disturbance levels described in Section~\ref{subsec:simulation_setting}. 
We set $d_{\mathrm{safe}}=200$\,m, corresponding to the side length of an air cube, which represents the exclusive airspace reserved for a vehicle in transit~\cite{muna2021air,jiang2022metrics}. 
For each combination of operational mode and disturbance level, we perform 30 simulation runs. 
We use the collision rate as the safety metric and the average number of vehicles that successfully exit the corridor network at $\CWP{3}$ without collision as the throughput metric.

\begin{table}[tbp]
\caption{Safety and throughput results under different disturbance levels and operational modes}
\begin{center}
\begin{tabular}{|c|c|c|c|}
\hline
\textbf{Dist. level} & \textbf{Mode} & \textbf{Collision [\%]} & \textbf{Successful exits} \\
\hline
\multirow{4}{*}{0} 
& Worst-case & 0.00 & 79.00 \\
& Stoc. $\sigma_a=6$\,m/s$^2$ & 0.00 & 87.00\\
& Stoc. $\sigma_a=3$\,m/s$^2$ & 0.00 & 107.00\\
& No ETA & 100.00 & 7.33 \\
\hline
\multirow{4}{*}{1} 
& Worst-case & 0.00 & 79.00 \\
& Stoc. $\sigma_a=6$\,m/s$^2$ & 0.00 & 87.00 \\
& Stoc. $\sigma_a=3$\,m/s$^2$ & 0.00 & 107.00 \\
& No ETA & 100.00 & 6.77 \\
\hline
\multirow{4}{*}{2} 
& Worst-case & 0.00 & 78.70\\
& Stoc. $\sigma_a=6$\,m/s$^2$ & 0.00 & 87.00 \\
& Stoc. $\sigma_a=3$\,m/s$^2$ & 0.00 & 107.00 \\
& No ETA & 100.00 & 5.87 \\
\hline
\multirow{4}{*}{3} 
& Worst-case & 0.00 & 78.50 \\
& Stoc. $\sigma_a=6$\,m/s$^2$ & 3.33 & 86.70\\
& Stoc. $\sigma_a=3$\,m/s$^2$ & 26.67 & 98.60 \\
& No ETA & 100.00 & 5.70 \\
\hline
\multirow{4}{*}{4} 
& Worst-case & 0.00 & 78.10 \\
& Stoc. $\sigma_a=6$\,m/s$^2$ & 66.67 & 61.03 \\
& Stoc. $\sigma_a=3$\,m/s$^2$ & 96.67 & 28.97 \\
& No ETA & 100.00 & 6.13 \\
\hline
\multirow{4}{*}{5} 
& Worst-case & 33.33 & 76.53 \\
& Stoc. $\sigma_a=6$\,m/s$^2$ & 93.33 & 50.70 \\
& Stoc. $\sigma_a=3$\,m/s$^2$ & 100.00 & 15.77 \\
& No ETA & 100.00 & 4.97 \\
\hline 
\end{tabular}
\label{tab:safety_throughput}
\end{center}
\end{table}

\begin{figure}[tbp]
\centering

\begin{subfigure}{0.85\columnwidth}
    \centering
    \includegraphics[width=\linewidth]{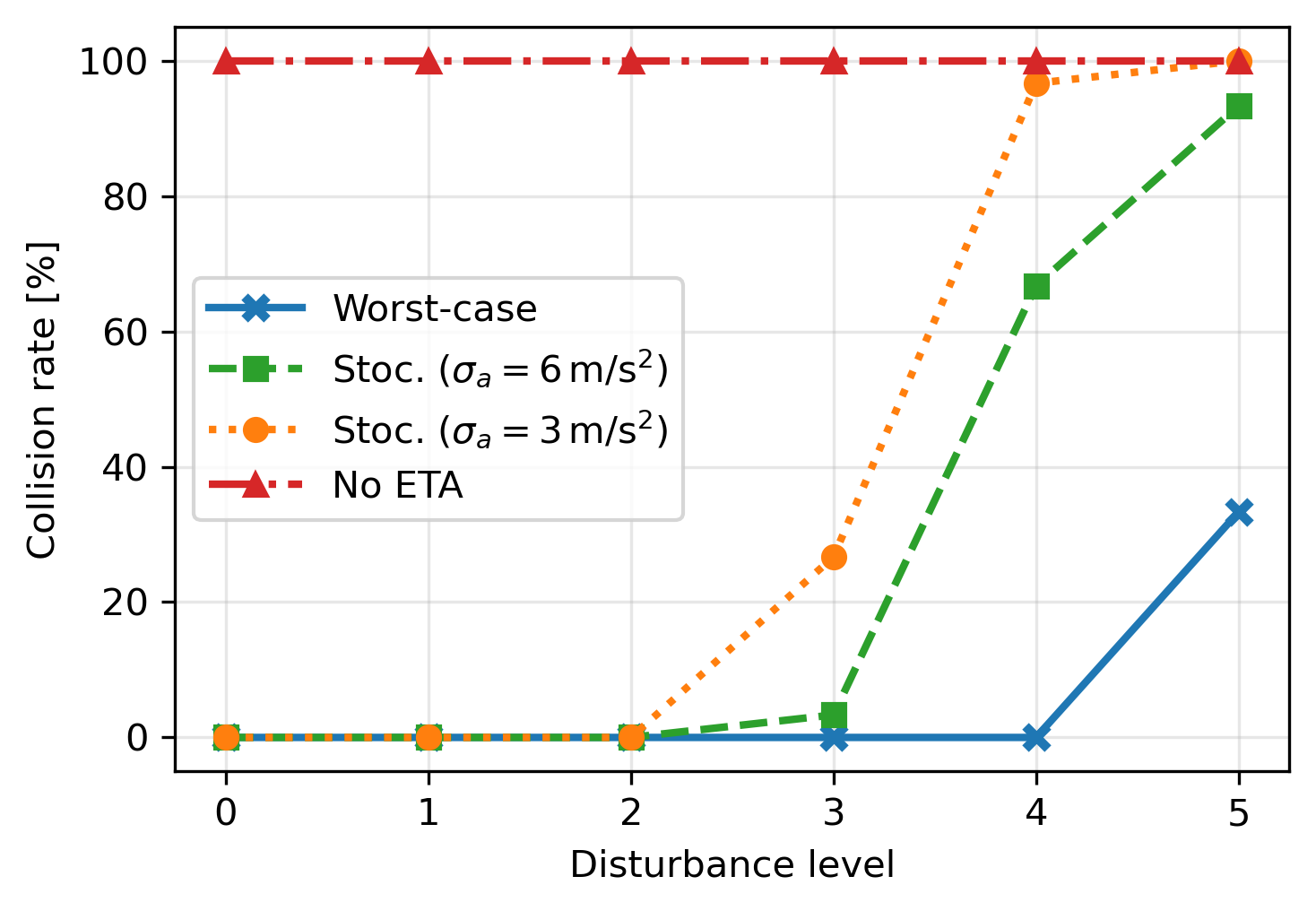}
    \caption{}
    \label{fig:collision_rate}
\end{subfigure}

\begin{subfigure}{0.85\columnwidth}
    \centering
    \includegraphics[width=\linewidth]{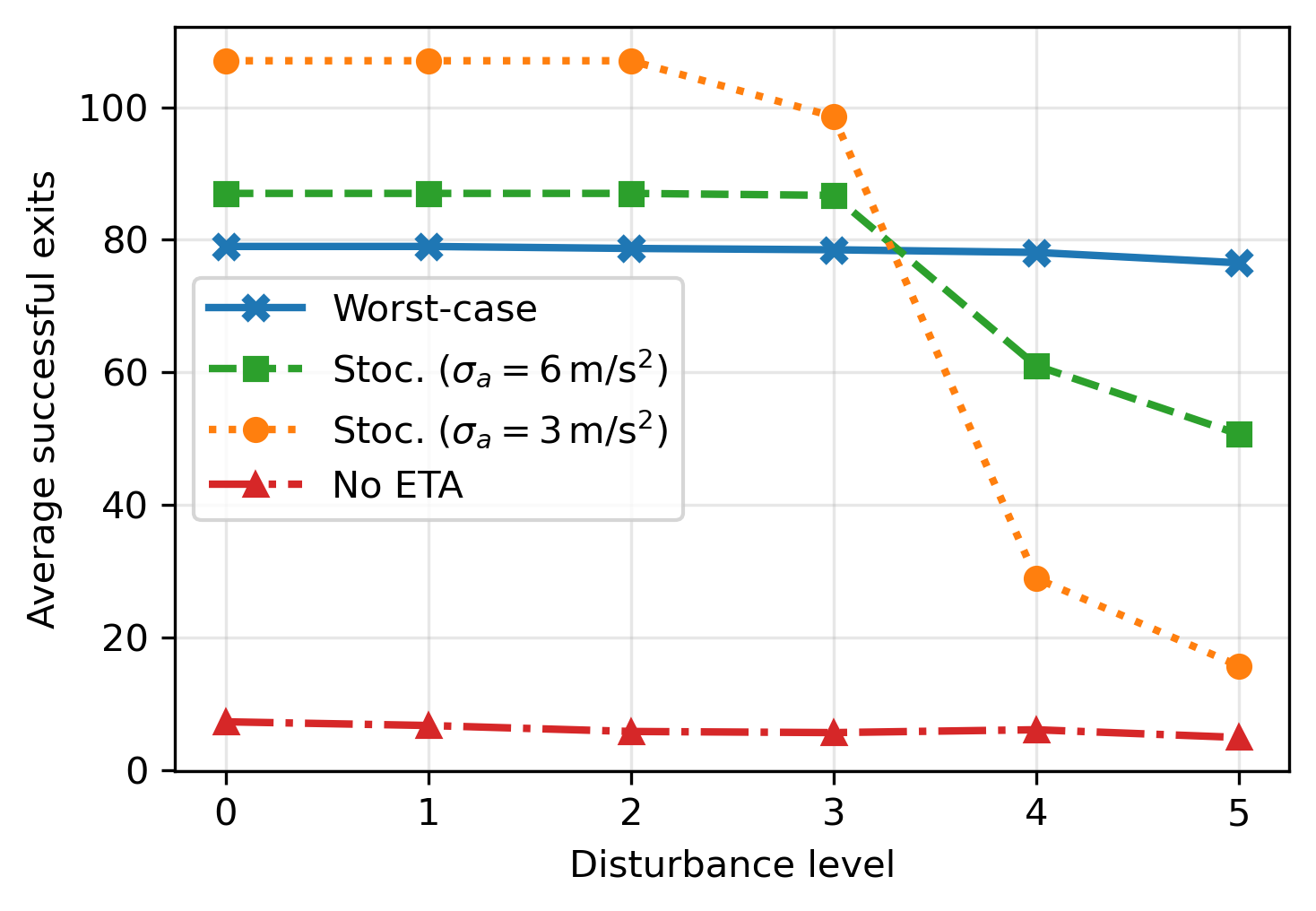}
    \caption{}
    \label{fig:successful_exits}
\end{subfigure}

\caption{Safety and throughput under different disturbance levels. %for the four operational modes. 
(a) Collision rate. 
(b) Average number of vehicles that successfully exit %the corridor network
at $\CWP{3}$ without collision.}
\label{fig:safety_throughput}
\end{figure}

Table~\ref{tab:safety_throughput} and Fig.~\ref{fig:safety_throughput} summarize the safety and throughput results, averaged over 30 simulation runs.
In Table~\ref{tab:safety_throughput}, the ``Successful exits'' column denotes the average number of vehicles that exit the corridor network at $\CWP{3}$ without collision.
Without ETA coordination, the collision rate is 100\% for all disturbance levels, and only about 5--7 vehicles successfully exit the corridor network without collision. 
This poor performance is expected because the No ETA baseline does not use CWP-level scheduling and assumes no coordination with other vehicles beyond a simple local rule: a vehicle brakes only when its distance to the preceding vehicle becomes no greater than $d_{\mathrm{safe}}+d_{\mathrm{margin}}$. 
In practical systems, vehicles may employ more advanced onboard sensing and collision-avoidance mechanisms. 
Here, the No ETA mode is used only as a simple baseline to highlight the benefit of adding ETA coordination at the scheduling level.  

Among the ETA-coordinated modes, the worst-case bound provides the strongest safety performance. 
It maintains zero collisions up to Disturbance Level~4 and reaches 33.33\% only at Level~5. 
The stochastic modes achieve higher throughput under mild disturbances, especially the $\sigma_a=3$\,m/s$^2$ mode, which yields 107 successful exits for Levels~0--2. 
However, this comes at the cost of reduced robustness under stronger disturbances: at Level~4, the collision rates of the $\sigma_a=6$\,m/s$^2$ and $\sigma_a=3$\,m/s$^2$ modes rise to 66.67\% and 96.67\%, respectively, and at Level~5 they reach 93.33\% and 100\%. 
These results illustrate the expected trade-off between throughput and robustness: less conservative bounds allow more vehicles to enter and exit under mild disturbances, whereas more conservative bounds provide stronger protection when vehicles are forced away from their scheduled ETAs.

% \subsection{Discussion on the Safety--Throughput Trade-off}
% \label{subsec:tradeoff_discussion}

% The results show a trade-off between safety robustness and throughput. 
% The worst-case bound gives the largest ETA gaps in Fig.~\ref{fig:computed_eta_gaps}, leading to the strongest safety performance in Table~\ref{tab:safety_throughput} and Fig.~\ref{fig:safety_throughput}, but also lower throughput. 
% The stochastic bounds allow smaller ETA gaps and therefore improve throughput under mild disturbances, but their robustness decreases when vehicles deviate strongly from the assumed stochastic model. 
% They also require more computation time for ETA-gap calculation, as discussed in Section~\ref{subsec:computed_eta_gaps}. 
% Using a larger uncertainty parameter, such as $\sigma_a=6$\,m/s$^2$, makes the stochastic bound more conservative and more robust than $\sigma_a=3$\,m/s$^2$, at the cost of reduced throughput.

% Overall, the stochastic bound may be suitable when strong disturbances are unlikely, and the distributions of vehicle speed and acceleration uncertainty are reasonably known. 
% In contrast, when such probabilistic assumptions are unreliable or higher levels of off-nominal disturbance are expected, the worst-case bound is a more conservative but safer choice.

\subsection{Discussion on the Safety--Throughput Trade-off}
\label{subsec:tradeoff_discussion}
The results show a trade-off between safety robustness and throughput. 
The worst-case bound gives the largest ETA gaps in Fig.~\ref{fig:computed_eta_gaps}, leading to the strongest safety performance in Table~\ref{tab:safety_throughput} and Fig.~\ref{fig:safety_throughput}, but also lower throughput under low disturbances.
The stochastic bounds allow smaller ETA gaps and thus improve throughput under mild disturbances, but their robustness decreases when vehicles deviate strongly from the assumed stochastic model. 
They also require more time for ETA-gap calculation, as discussed in Section~\ref{subsec:computed_eta_gaps}. 
Using a larger uncertainty parameter, such as $\sigma_a=6$\,m/s$^2$, makes the stochastic bound more conservative and more robust than $\sigma_a=3$\,m/s$^2$, at the cost of reduced throughput. 
Overall, the stochastic bound may be suitable when strong disturbances are unlikely and the uncertainty distributions are reasonably known. 
In contrast, when such probabilistic assumptions are unreliable or stronger off-nominal disturbances are expected, the worst-case bound is the more conservative but safer choice.

These results also show a limitation of the present ETA-based framework, which assumes that vehicles can follow the planned ETA sequence while remaining within the corridor speed constraints. 
Under sufficiently strong disturbances, this assumption may fail, and vehicles may no longer be able to realize the scheduled ETAs. 
In this sense, the disturbance model reveals the robustness limit of the framework. 
Nevertheless, the results show that ETA-based coordination remains effective over a meaningful range of disturbance levels, with the robustness margin depending on the selected trajectory bound. 
Developing tactical-phase mechanisms for predicting and updating ETAs at CWPs under delay is therefore an important direction for future work.

% These results also highlight a limitation of the present ETA-based framework. 
% The method assumes that vehicles can follow the planned ETA sequence while remaining within the corridor speed constraints. 
% Under sufficiently strong disturbances, this assumption may fail, and vehicles may no longer be able to realize the scheduled ETAs. 
% In this sense, the disturbance model reveals the robustness limit of the framework. 
% Nevertheless, the results show that ETA-based coordination remains effective over a meaningful range of disturbance levels, with the robustness margin depending on the selected safety bound. 
% It is therefore an important direction for future work to develop tactical-phase mechanisms for predicting and updating ETAs at CWPs when delays occur.

% The framework provides a lightweight way to support corridor management using CWP-level timing rather than full trajectory optimization.

\section{Conclusion}
\label{sec:conclusion}

This paper proposed an ETA-based coordination framework for safety-aware scheduling in merging UAM corridors. 
The framework computes sufficient ETA gaps at the merging point by considering the corridor sections shared by consecutive vehicles. 
We developed two types of trajectory bounds: a worst-case spatio-temporal bound based on extreme admissible speed profiles, and a stochastic bound based on probabilistic position envelopes under acceleration uncertainty.

Numerical simulations showed the trade-off between safety robustness and throughput. 
The worst-case bound provides stronger robustness when disturbance levels are high, but results in larger ETA gaps and lower throughput.
The stochastic bound can improve throughput under mild disturbances, but its performance depends on the assumed uncertainty model.

% Future work will investigate methods to improve robustness under high disturbances, where the current stochastic assumptions may no longer be reliable. 
% We will also study faster stochastic ETA-gap computation. 
% Another important direction is to develop corridor-design strategies that further optimize the balance between safety and efficiency in merging UAM operations. 
% We also plan to extend the framework to tactical-phase operations, which will require online mechanisms for ETA prediction and updating at CWPs under delays.

One promising future direction is to adaptively combine the worst-case and stochastic bounds, for example, by switching ETA-gap computation or adjusting the stochastic uncertainty level based on measured ETA deviations, speed deviations, or disturbance severity.
Such an approach could improve throughput under mild disturbances while preserving robustness when stochastic assumptions become unreliable.
Future work will also investigate faster stochastic ETA-gap computation, corridor-design strategies that further optimize the balance between safety and efficiency in merging UAM operations, and tactical-phase extensions with online ETA prediction and updating at CWPs under delays.

\section*{Acknowledgment}
This work is based on results obtained from a project, JPNP22002, commissioned by the New Energy and Industrial Technology Development Organization (NEDO).

During the preparation of this work, the authors used ChatGPT-5.4 and Grammarly to support grammar and spelling checks, paraphrasing, and wording refinement. 
The authors reviewed and edited the content after using these tools and are fully responsible for the final content of the publication.

% \section*{Acknowledgment}

% The preferred spelling of the word ``acknowledgment'' in America is without 
% an ``e'' after the ``g''. Avoid the stilted expression ``one of us (R. B. 
% G.) thanks $\ldots$''. Instead, try ``R. B. G. thanks$\ldots$''. Put sponsor 
% acknowledgments in the unnumbered footnote on the first page.

% \section*{References}
\bibliographystyle{IEEEtran}
\bibliography{ref}

\end{document}